\RequirePackage{fixltx2e}

\documentclass[aps,pre,reprint,groupedaddress,nofootinbib,floatfix,final]{revtex4-1}

  \usepackage[pdftex]{graphicx}
  \graphicspath{{}}
  \DeclareGraphicsExtensions{.pdf}
  \usepackage[abs]{overpic}
\usepackage[cmex10]{amsmath}
\usepackage{amsfonts}
\usepackage{mathtools}
\usepackage{mathrsfs}
\usepackage[ampersand]{easylist}

\usepackage{array}
\usepackage{tabularx,booktabs}
\usepackage{mdwmath}
\usepackage[caption=false,font=footnotesize]{subfig}
\usepackage{afterpage}
\usepackage{fixltx2e}
\usepackage{dblfloatfix}
\usepackage{float}
\usepackage{placeins}
\usepackage{newfloat}
\DeclareFloatingEnvironment[name={S}]{suppfigure}

\usepackage{url}
\begin{document}


\title{Observability and Controllability of Nonlinear Networks: \\The Role of Symmetry}


\author{Andrew J. Whalen}
\email[]{awhalen@psu.edu}
\affiliation{Department of Mechanical Engineering}
\affiliation{Center for Neural Engineering, The Pennsylvania State University, University Park, PA 16802 }

\author{Timothy D. Sauer}
\email[]{tsauer@gmu.edu}
\affiliation{Department of Mathematical Sciences, George Mason University, Fairfax, VA 22030 }

\author{Sean N. Brennan}
\email[]{sbrennan@psu.edu}
\affiliation{Department of Mechanical Engineering}
\affiliation{Center for Neural Engineering, The Pennsylvania State University, University Park, PA 16802 }

\author{Steven J. Schiff}
\email[]{sschiff@psu.edu}
\affiliation{Center for Neural Engineering, Departments of Engineering Science and Mechanics, Neurosurgery and Physics, The Pennsylvania State University, University Park, PA 16802 }


\date{\today}

\begin{abstract}
Observability and controllability are essential concepts to the design of predictive observer models and feedback controllers of networked systems. For example, noncontrollable mathematical models of real systems have subspaces that influence model behavior, but cannot be controlled by an input. Such subspaces can be difficult to determine in complex nonlinear networks. Since almost all of the present theory was developed for linear networks without symmetries, here we present a numerical and group representational framework, to quantify the observability and controllability of nonlinear networks with explicit symmetries that shows the connection between symmetries and nonlinear measures of observability and controllability. We numerically observe and theoretically predict that not all symmetries have the same effect on network observation and control. Our analysis shows that the presence of symmetry in a network may decrease observability and controllability, although networks containing only rotational symmetries remain controllable and observable. These results alter our view of the nature of observability and controllability in complex networks, change our understanding of structural controllability, and affect the design of mathematical models to observe and control such networks.
\end{abstract}


\maketitle

\section{Introduction}
An observer model of a natural system has many useful applications in science and engineering, including understanding and predicting weather or controlling dynamics from robotics to neuronal systems \cite{Schiff2012}. A fundamental question that arises when utilizing filters to estimate the future states of a system is how to choose a model and measurement function that faithfully captures the system dynamics and can predict future states \cite{Voss2004,Sauer2009}. An observer is a model of a system or process that assimilates data from the natural system being modeled \cite{Kalnay2003}, and reconstructs unmeasured or inaccessible variables. In linear systems, the key concept to employ a well designed observer is observability, which quantifies whether there is sufficient information contained in the measurement to adequately reconstruct the full system dynamics \cite{Kalman1963,Luenberger1971}.

An important problem when studying networks is how best to observe and control the entire network when only limited observation and control input nodes are available. In classic work, Lin \cite{Lin1974} described the topologies of graph directed linear networks that were structurally controllable. Incorporating Lin's framework, Liu et al \cite{Liu2011} described an efficient strategy to count the number of control points required for a complex network, which have an interesting dependence on time constant \cite{Cowan2012}. Structural observability is dual to structural controllability \cite{Rech1990}. In \cite{Liu2012}, the requirements of structural observability incorporated explicit use of transitive components of directed graphs - fully connected subgraphs where paths lead from any node to any other node - to identify the minimal number of sites required to observe from a network.

All of these prior works depend critically on the dynamics being linear and generic, in the sense that network connections are essentially random. Joly \cite{Joly2012} showed that transitive generic networks with nonlinear nodal dynamics are observable from any node. Nevertheless, symmetries are present in natural networks, as evident from their known structures \cite{Weyl1952} as well as the presence of synchrony. Recently, Golubitsky et al \cite{Golubitsky2012} proved the rigid phase conjecture - that the presence of synchrony in networks implies the presence of symmetries and vice versa. In particular, synchrony is an intrinsic component of brain dynamics in normal and pathological brain dynamics \cite{Uhlhaas2006}.

Our present work is motivated by the question: what role do the symmetries and network coupling strengths play when reconstructing or controlling network dynamics? The intuition here is straightforward: consider 3 linear systems with identical dynamics (diagonal terms of the system matrix \(A \text{ in } \dot{\mathbf{x}}(t)=A\mathbf{x}(t)\)), if the coupling terms are identical (off-diagonal terms of \(A\)), it is easy to show that the resulting observability of individual states becomes degenerate as the rows and columns of the system matrix become linearly dependent under elementary matrix operations. For example, consider the trivial case of a 3x3 system matrix of ones:
\begin{equation}
	\dot{\mathbf{x}}=A\mathbf{x}=
	\begin{bmatrix}
		1 & 1 & 1 \\
		1 & 1 & 1 \\
		1 & 1 & 1 \\
	\end{bmatrix} 
	\begin{bmatrix}
		x_1 \\
		x_2 \\ 
		x_3 
	\end{bmatrix}.
\end{equation}
The system is degenerate in the sense that there is only one dynamic, as the rows and columns of A are not independent. This lack of independent rows and columns of the system matrix has direct implications for the controllability and observability of the system. For example, in this trivial system the difference between any two of the states is constrained to a constant \(x_1-x_2=c\), thus there is no input coupled to the third state \(x_3\) that could control both \(x_1\) and \(x_2\) independently from each other. 

In fact, for the more general case of linear time-varying networks, group representation theory \cite{Burnside1955} has been utilized to show that linear time-varying networks can be non-controllable or non-observable due to the presence of symmetry in the network \cite{Rubin1972}. Brought into context, in networks with symmetry Rubin \& Meadows \cite{Rubin1972} defines a coordinate transform which decomposes the network into decoupled observable (controllable) and unobservable (uncontrollable) subspaces, which then can be determined by inspection like our previous trivial example. Recently, Pecora et al \cite{Pecora2014} utilized this same method to show how separate subsets of complex networks could sychronize and desychronize according to these same symmetry-defined subspaces. Interestingly, while \cite{Rubin1972} has been a rather obscure work, it is based on Wigner's work in the 1930's applying group representation theory to the mechanics of atomic spectra \cite{Wigner1959}. Thus, just as the structural symmetry of the Hamiltonian can be used to simplify the solution to the Schr\"{o}dinger equation \cite{Tinkham1964}, the topology of the coupling in a network can have a profound impact on its observation and control. 

In this article, we extend the exploration of observability and controllability to network motifs with explicit nonlinearities and symmetries. We further explore the effect of coupling strength within such networks, as well as spatial and temporal effects on observability and controllability. Lastly, we demonstrate the utility of the linear analysis of group representation theory as a tool with which to gain insights into the effects of symmetry in nonlinear networks.

\section{Background}
From the theories of differential embeddings \cite{Whitney1936} and nonlinear reconstruction \cite{Takens1981, Sauer1991} we can create a nonlinear measure of observability comprised of a measurement function and its higher Lie derivatives employing the differential embedding map \cite{Letellier2005}. The differential embedding map of an observer provides the information contained in a given measurement function and model, which can be quantified by an index \cite{Friedland1975,Gibson1992}. Computed from the Jacobian of the differential embedding map, the observability index is a matrix condition number which quantifies the perturbation sensitivity (closeness to singularity) of the mapping created by the measurement function used to observe the system. There is a dual theory for controllability, where the differential embedding map is constructed from the control input function and its higher Lie brackets with respect to the nonlinear model function \cite{Haynes1970,Hermann1977}. Singularities in the map cause information about the system to be lost and observability to decrease. Additionally, the presence of symmetries in the system's differential equations makes observation difficult from variables around which the invariance of the symmetry is manifested \cite{Letellier2002,Pecora1990}. We extend this analysis to networks of ordinary differential equations and investigate the effects of symmetries on observability and controllability of such networks as a function of connection topology, measurement function, and connection strength. 

\subsection{Linear Observability and Controllability}
In the early 1960s, Rudolph Kalman introduced the notions of state space decomposition, controllability and observability into the theory of linear systems \cite{Kalman1963}. From this work comes the classic concept of observability for a linear time-invariant (LTI) dynamic system, which defines a `yes' or `no' answer whether a state can be reconstructed from a measurement using a rank condition check.

A dynamic model for a linear (time-invariant) system can be represented by
\begin{equation}
\label{LTIsys}
	\begin{aligned}
		 &\dot{\mathbf{x}}(t) = A \mathbf{x}(t)+B \mathbf{u}(t) \\
		&y(t) = C \mathbf{x}(t),
	\end{aligned}
\end{equation}
where \( \mathbf{x}\in \mathbb{R}^n \) represents the state variable, \( \mathbf{u}\in \mathbb{R}^m \) is the external input to the system and \( y\in \mathbb{R}^p \) is the output (measurement) function of the state variable. Typically there are less measurements than states, so \(p<n\).  The intuition for observability comes from asking whether an initial condition can be determined from a finite period of measuring the system dynamics from one or more sensors. That is, given the system in (\ref{LTIsys}), with \(\mathbf{x}(t)=e^{At}\mathbf{x_{0}}\) and \(B\mathbf{u}=0\), determine the initial condition \(\mathbf{x_{0}}\) from measurement \(y(t), 0\leq t \leq T\). To evaluate this locally, we take the higher derivatives of \(y(t)\): 
\begin{equation}
\label{yDerivs}
	\begin{aligned}
		&y(t) = C \mathbf{x}(t) \\
		&\dot{y}=C\dot{\mathbf{x}}(t)=CA\mathbf{x}(t) \\
		&\ddot{y}=CA\dot{\mathbf{x}}(t)=CA^2\mathbf{x}(t) \\
		&\vdots \\
		&y^{(n-1)}=CA^{n-1}\mathbf{x}(t).
	\end{aligned}
\end{equation}
Factoring the \(\mathbf{x}\) terms and putting \(y\) and its higher derivatives in matrix form, we have a mapping from outputs to states
\begin{equation}
\label{yOx}
	\begin{bmatrix}
		y\\
		\dot{y}\\
		\ddot{y}\\
		\vdots\\
		y^{(n-1)}		
	\end{bmatrix}
	=
	\begin{bmatrix}
		C\\
		CA\\
		CA^2\\
		\vdots\\
		CA^{n-1}
	\end{bmatrix}
\mathbf{x},
\end{equation}
where the linear observability matrix \cite{Kailath1980}  is defined as
\begin{equation}
\label{ObsMatrix}
	O\equiv
	\begin{bmatrix}
		C\\
		CA\\
		CA^2\\
		\vdots\\
		CA^{n-1}
	\end{bmatrix}
\end{equation}
The finite limit of taking derivatives in (\ref{yDerivs}) comes from the Cayley-Hamilton theorem, which specifies that any square matrix A satisfies is own characteristic equation, which is the polynomial \(p(\lambda)=0\) where \(p(\lambda)=(\lambda I_{n} - A)\). In other words, \(A^n\) is spanned by the lower powers of \(A\), from \(A^0\) to \(A^{n-1}\),
\begin{equation}
	\begin{aligned}
		y(t)=C& e^{At}\mathbf{x_{0}}, \quad \text{with} \quad e^{At}\equiv\sum_{k=0}^{n-1}\alpha_k(t)A^k \\
		y(t)=[&\alpha_0(t)C+\alpha_1(t)CA+\alpha_2(t)CA^2+ \\ &\hdots+\alpha_{n-1}(t)CA^{n-1}]\mathbf{x_0}.
	\end{aligned}
\end{equation}
Thus, if the observability matrix spans \(n\) space (rank(\(O\))\(=n\)), the initial condition \(x_0\) can be determined, as the mapping \(x_0=O^{-1}y(t)\) from output to states exists and is unique. More formally, the system (\ref{LTIsys}) is locally observable (distinguishable at a point \(x_0\)) if there exists a neighborhood of \(x_0\) such that \(x_0 \neq x_1 \implies y(x_0) \neq y(x_1)\).

In a similar fashion, the linear controllability matrix is derived from asking whether an input \(\mathbf{u}(t)\) can be found to take any initial condition \(\mathbf{x}(0)=x_0\) to arbitrary position \(\mathbf{x}(T)=x_f\) in a finite period of time \(T\). For the sake of simplicity, we assume a single input \(u(t)\) and take the higher derivatives of \(\dot{\mathbf{x}(t)}=A\mathbf{x}(t)+Bu(t)\) up to the \((n-1)^{th}\) derivative of \(u(t)\) (again using the Cayley-Hamilton theorem):
\begin{equation}
\label{xDerivs}
	\begin{aligned}
		&\dot{\mathbf{x}(t)} = A \mathbf{x}(t)+Bu(t) \\
		&\ddot{\mathbf{x}(t)}=A^2\mathbf{x}(t)+ABu(t)+B\dot{u}(t) \\
		&\dddot{\mathbf{x}(t)}=A^3\mathbf{x}(t)+A^2Bu(t)+AB\dot{u}(t)+B\ddot{u}(t) \\
		&\vdots \\
		&\mathbf{x}^{(n)}(t)=A^{n}\mathbf{x}(t)+A^{n-1}Bu(t)+A^{n-2}B\dot{u}(t)+\\
		&\qquad\hdots+Bu^{(n-1)}(t)
	\end{aligned}
\end{equation}
which gives us a mapping from input to states
\begin{equation}
\label{xn-AQu}
	\begin{bmatrix}
		\dot{x}(t)\\
		\ddot{x}(t)\\
		\vdots\\
		x^{(n-1)}(t)\\
		x^{(n)}(t)		
	\end{bmatrix}-
	\begin{bmatrix}
		A\\
		A^2\\
		\vdots\\
		A^{(n-1)}\\
		A^{(n)}		
	\end{bmatrix}A^n\mathbf{x}(t)		
	=
	Q
	\begin{bmatrix}
		u(t)\\
		\dot{u}(t)\\
		\vdots\\
		u^{(n-2)}(t)\\
		u^{(n-1)}(t)\\		
	\end{bmatrix}
\end{equation}
where the linear controllability matrix is defined \cite{Kailath1980} as
 \begin{equation}
\label{CtrlbMatrix}
	Q\equiv
	\begin{bmatrix}
		B,AB,A^2B\hdots,A^{n-1}B
	\end{bmatrix}.
\end{equation}

\subsection{Differential Embeddings and Nonlinear Observability}
From early work on the nonlinear extensions of observability in the 1970s \cite{Haynes1970,Hermann1977}, it was shown that the observability matrix for nonlinear systems could be expressed using the measurement function and its higher order Lie derivatives with respect to the nonlinear system equations. The core idea is to evaluate a mapping \(\phi \) from the measurements to the states \(\phi:\mathbb{R}^p \xrightarrow{} \mathbb{R}^n\). In particular, Hermann and Krener \cite{Hermann1977} showed that the space of the measurement function is embedded in \( \mathbb{R}^n \) when the mapping from measurement to states is everywhere differentiable and injective by the Whitney Embedding Theorem \cite{Whitney1936,Takens1981}. An embedding is a map involving differential structure that does not collapse points or tangent directions \cite{Sauer1991}, thus a map \( \phi \) is an embedding when the determinant of the map Jacobian \(\text{Det}(\frac{\partial\phi}{\partial \mathbf{x}}|_{\forall \mathbf{x} \in \mathbb{R}^n}) \) is non-vanishing and one-to-one (injective). In a recent series of papers \cite{Letellier1998,Letellier2002,Letellier2005}, Letellier et al. computed the nonlinear observability matrices for the well-known Lorenz and R\"{o}ssler systems \cite{Lorenz1963,Rossler1976} and demonstrated that the order of the singularities present in the observability matrix (and thus the amount of intersection between the singularities and the phase space trajectories) was related to the decrease in observability. It is worth noting that the calculation of the observability matrix and locally evaluating the conditioning of the matrix over a state trajectory is a straightforward process and much more tractable than analytically determining the singularities (and thus their order) of the observability matrix of a system of arbitrary order. The former is limited only by computational capacity and the differentiability of the system equations to order \(n-1\), where \(n\) is the order of the system.

For a nonlinear system, we replace \(A\mathbf{x}(t)\) in (\ref{LTIsys}) by a nonlinear vector field \(A_{NL}(\mathbf{x}(t))\), and assume that the smooth scalar measurement function is taken as \( y(t)=C\mathbf{x}(t)\) and the system equations comprise the nonlinear vector field \(\mathbf{f}(\mathbf{x}(t))=A_{NL}(\mathbf{x}(t)\)) (note: if there is no external input, then \(B\mathbf{u}(t)=0\) which we assume here to simplify the display of equations\footnote{If \(B\mathbf{u}\neq0\) then as long as the input is known the mapping from output to states can be solved, and the determination of observability still relies on the conditioning of the matrix \(O\).}). As in the linear case, we evaluate locally by taking the higher Lie derivatives of \( y(t)\), and for compactness of notation dependence on \(t\) is implied:
\begin{equation}
\label{LieDeriv}
	\begin{aligned}
		\mathfrak{L}^{0}_{f}(y&(x))= y(x) \\
		\mathfrak{L}^{1}_{f}(y&(x))=\nabla y(x)\cdot\mathbf{f}(x)= \frac{\partial y(x)}{\partial x} \cdot \mathbf{f}(x) \\
		\mathfrak{L}^{2}_{f}(y&(x))= \frac{\partial}{\partial x}[\mathfrak{L}^{1}_{f}(y(x))] \cdot \mathbf{f}(x) \\
		&\vdots \\
		\mathfrak{L}^{k}_{f}(y&(x))= \frac{\partial}{\partial x}[\mathfrak{L}^{k-1}_{f}(y(x))] \cdot \mathbf{f}(x)
	\end{aligned}
\end{equation}
where \(\mathfrak{L}_{f}(y(x))\) is the Lie derivative of \(y(x)\) along the vector field \(\mathbf{f}(x)\). More explicitly, we have \(\mathbf{x}\in \mathbb{R}^n\), so as a vector example the first Lie derivative will take the form
\begin{equation}
\label{LieDeriv2}
	\mathfrak{L}^{1}_{f}(y(x))= 
	\begin{bmatrix}
		\frac{\partial y(x)}{\partial x_1} \hdots \frac{\partial y(x)}{\partial x_n}
	\end{bmatrix}
	\cdot
	\begin{bmatrix}
		f_1(\mathbf{x}) \\
		\vdots \\
		f_n(\mathbf{x})
	\end{bmatrix}.
\end{equation}
With formal definitions of the measurement (output) function (\ref{LTIsys}) and its higher Lie derivatives (\ref{LieDeriv}), the differential embedding map \(\phi\) is defined as the Lie derivatives \(\mathfrak{L}^{0}_{f}(y(x)) \hdots \mathfrak{L}^{n-1}_{f}(y(x))\), where the superscripts represent the order of the Lie derivative from \(0 \text{ to } n-1\), where \(n\) is the order of the system \(A_{NL}(\mathbf{x})\)
\begin{equation}
\label{phi}
	\phi = 
	\begin{bmatrix}
		\mathfrak{L}^{0}_{f}(y(x)) \\[0.3em]
		\mathfrak{L}^{1}_{f}(y(x)) \\
		\vdots \\
		\mathfrak{L}^{n-1}_{f}(y(x))
	\end{bmatrix}.
\end{equation}
Taking the Jacobian of the map \(\phi\) we arrive at the observability matrix
\begin{equation}
\label{ObsMatrixNL}
	O\equiv\frac{\partial\phi}{\partial\mathbf{x}}=
	\begin{bmatrix}
		\frac{\partial\mathfrak{L}^{0}_{f}(y(x))}{\partial\mathbf{x_1}} & \hdots & \frac{\partial\mathfrak{L}^{0}_{f}(y(x))}{\partial\mathbf{x_n}}\\[0.3em]
		\vdots & \ddots & \vdots \\
		\frac{\partial\mathfrak{L}^{n-1}_{f}(y(x))}{\partial\mathbf{x_1}} & \hdots & \frac{\partial\mathfrak{L}^{n-1}_{f}(y(x))}{\partial\mathbf{x_n}}
	\end{bmatrix},
\end{equation}
which reduces to (\ref{ObsMatrix}) for linear system representations. The key intuition here is that in the nonlinear case the observability matrix becomes a function of the states, where a linear system is always a constant matrix of parameters.
 

\subsection{Lie Brackets and Nonlinear Controllability}
The nonlinear controllability matrix is developed in \cite{Haynes1970} from intuitive control problem examples and given rigorous treatment in \cite{Hermann1977}; in a dual fashion to observability, the controllability matrix is a mapping constructed from the input function and its higher order Lie brackets. The Lie bracket is an algebraic operation on two vector fields \(\mathbf{f}(x),\mathbf{g}(x) \in \mathbb{R}^n\) that creates a third vector field \(\mathfrak{F}(x)\), which when taken with \(\mathbf{g}\) as the input control vector \(\mathbf{u} \in \mathbb{R}^m\) defines an embedding in \(\mathbb{R}^n\) that maps the input to states \cite{Hermann1977}. 

For a nonlinear system, we replace \(A\mathbf{x}(t)\) in (\ref{LTIsys}) by a nonlinear vector field \(A_{NL}(\mathbf{x}(t))\), take the input function as \(g=B\mathbf{u}(t)\) in system (\ref{LTIsys}), and create Lie brackets with respect to the nonlinear vector field \(\mathbf{f}(\mathbf{x}(t))=A_{NL}(\mathbf{x}(t))\). The Lie bracket is defined as
\begin{equation}
\label{LieBracket}
	\begin{aligned}
		(ad^{1}_{\mathbf{f}}&,g)=[\mathbf{f},\mathbf{g}]=\frac{\partial\mathbf{g}}{\partial\mathbf{x}}\mathbf{f}-\frac{\partial\mathbf{f}}{\partial\mathbf{x}}\mathbf{g} \\
		(ad^{2}_{\mathbf{f}}&,g) =[\mathbf{f},[\mathbf{f},\mathbf{g}]]=\frac{\partial(ad^{1}_{\mathbf{f}},g)}{\partial\mathbf{x}}\mathbf{f}-\frac{\partial\mathbf{f}}{\partial\mathbf{x}}(ad^{1}_{\mathbf{f}},g)\\
		&\vdots \\
		(ad^{k}_{\mathbf{f}}&,g)=[\mathbf{f},(ad^{k-1}_{\mathbf{f}},\mathbf{g})],
	\end{aligned}
\end{equation}
where \((ad^{k}_{\mathbf{f}},g)\) is the adjoint operator and the superscripts represent the order of the Lie bracket. With formal definitions of the input function (\ref{LTIsys}) and its higher Lie brackets (\ref{LieBracket}) from \(1 \text{ to } n\), where \(n\) is the order of the system matrix \(A_{NL}(\mathbf{x}(t))\), the nonlinear controllability matrix is defined as 
\begin{equation}
\label{CtrlbMatrixNL}
	\begin{aligned}
		Q\equiv&
		\begin{bmatrix}
			\mathbf{g},(ad^{1}_{\mathbf{f}},g),\hdots,(ad^{n}_{\mathbf{f}},g)
		\end{bmatrix} \\
		=&
		\begin{bmatrix}
			\mathbf{g},[\mathbf{f},\mathbf{g}],[\mathbf{f}[\mathbf{f},\mathbf{g}]],\hdots,[\mathbf{f},(ad^{n-1}_{\mathbf{f}},\mathbf{g})]
		\end{bmatrix}.
	\end{aligned}
\end{equation}

\subsection{Observability/Controllability Index}
In systems with real numbers, calculation of the Kalman rank condition may not yield an accurate measure of the relative closeness to singularity (conditioning) of the observability matrix. It was demonstrated in \cite{Friedland1975} that the calculation of a matrix condition number \cite{Strang2005} would provide a more robust determination of the ill-conditioning inherent in a given observability matrix, since condition number is independent of scaling and is a continuous function of system parameters (and states in the generic nonlinear case). We will use the inverted form of the observability index \( \delta(\mathbf{x})\) given in \cite{Friedland1975} so that \(0\leq\delta(\mathbf{x})\leq1 \)
\begin{equation}
\label{ObsIndex}
	\delta(\mathbf{x})=\frac{|\sigma_{min}[O^TO]|}{|\sigma_{max}[O^TO]|},
\end{equation} 
where \(\sigma_{min} \text{ and } \sigma_{max}\) are the minimum and maximum singular values of \(O^TO\) respectively and \( \delta(\mathbf{x})=1 \) indicates full observability while \( \delta(\mathbf{x})=0 \) indicates no observability \cite{Aguirre1995}. Similarly the controllability index is just (\ref{ObsIndex}) with the substitution of \(Q\) for \(O\).

\section{Observability and Controllabilty of 3-node Fitzhugh-Nagumo network motifs}
\subsection{Fitzhugh-Nagumo System Dynamics}
The Fitzhugh-Nagumo (FN) equations \cite{Fitzhugh1961,Nagumo1962}, comprise a general representation of excitable neuronal membrane. The model is a 2-dimensional analogue of the well known Hodgkin-Huxley model \cite{Hodgkin1952} of an axonal excitable membrane. The nonlinear FN model can exhibit a variety of dynamical modes which include active transients, limit cycles, relaxation oscillations with multiple time scales, and chaos \cite{Fitzhugh1961,Doi1995}. A nonlinear connection function will be used to emulate properties of neuronal synapses. 

The system dynamics at a node are given by the (local 2\textsuperscript{nd} order) state space 
\begin{equation}
\label{FNsys}
	\begin{aligned} 
		&\dot{v_i} = c (v_i-\frac{v_i^3}{3}-w_i+ \sum f_{NL}(v_{j},d_{ij})+I) \\
		&\dot{w_i} = v_i-bw_i+a,
	\end{aligned} 
\end{equation}
where \(i=1,2,3\) for the 3-node system, \( v_i \) represents membrane voltage of node \(i\), \( w_i \) is recovery, \( d_{ij}\) the inter-nodal distance from node \( j\) to \(i\), \(v_j\) the voltage of neighbor nodes with \(j=1,2,3\) and \(j \neq i\),  input current \(I\), and the system parameters \(a=0.7, b=0.8, c=10\). As defined above in Eqns. (\ref{ObsMatrixNL}) and (\ref{CtrlbMatrixNL}), the observability and controllability matrices are a function of the states which means a dependence on the particular trajectory taken in phase space. In the following analysis, we are interested in directed information flow between nodes as a function of various topological connection motifs, connection strengths and input forcing functions (which provide different trajectories through phase space). Each motif is representative of a unique combination of directed connections between the 3 nodes with and without latent symmetries. The nonlinear connection function commonly used in neuronal modeling \cite{Koch2003} takes the form of the sigmoidal activation function of neighboring activity (a hyperbolic tangent) and an exponential decay with inter-nodal distance. We utilize various coupling strengths to determine the effects on the observability (controllability) of the network. Our coupling function takes the form
\begin{equation}
\label{FNconnection}
	f_{NL}(v,d)=\frac{k}{2}(\operatorname{tanh}(\frac{v-h}{2m})+1)e^{-d}.
\end{equation}
The sigmoid parameters \(k=1, h=0, m=1/4\), are set such that \(f_{NL}(v,d)\) has an output range \([0,1]\) for the input interval \([-2,2]\), which is the range of the typical FN voltage variable. To introduce heterogeneity for symmetry breaking a \(10\%\) variance noise term was added to each of the \(d_{ij}\) terms (there are 6 total possible coupling terms \(d_{12},d_{13}\dots\) etc.). 

In this configuration, inputs from neighboring nodes act in an excitatory-only manner, while the driving input current was a square wave \(I=0.25 [ \sum_{n=-\infty}^{\infty} \sqcap (\omega t-n T)+1]\) (where \(\sqcap\) is the rectangular function, \(\omega=2\pi/5\) and \(T=16 \frac{2}{3}\)) applied to all three nodes to provide a limit cycle regime to the network; for the limit cycle regime generated in the original paper by Fitzhugh \cite{Fitzhugh1961}, the driving current input was constant \( I=-0.45\) (with the system parameters mentioned above) which we will also explore. Chaotic dynamics were generated with a slightly different square wave input \cite{Doi1995} \(I=0.1225 [ \sum_{n=-\infty}^{\infty} \sqcap (\omega t-n T)+1]\) (with \(\omega=2\pi/1.23\) and \(T=2.7891\)) also applied to all three nodes. These various driving input regimes allow a wider exploration of the phase space of the system as each driving input commands a different trajectory, which will in turn influence the observability and controllability matrices.

\begin{figure}[!ht]
\includegraphics[width=3.3in, clip=true, trim = 110 290 110 20]{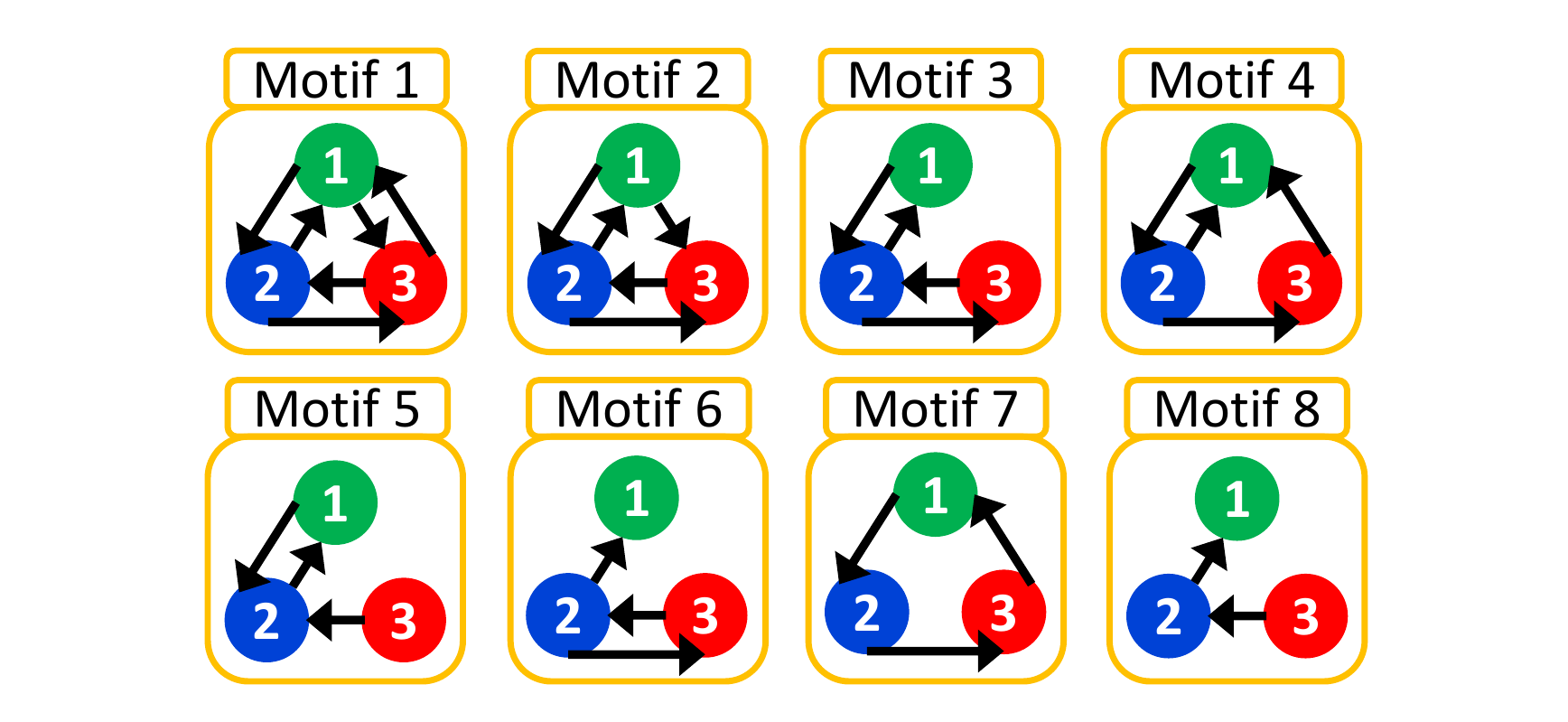}
\caption{\label{Motifs}The eight different 3-node network connection motifs studied.}
\end{figure}

\subsection{Network Motifs and Simulated Data}
As we are interested in the effect of connection topology on observability and controllability, we study the simplest nontrivial network: a 3-node network. Such small network motifs are highly overrepresented in neuronal networks \cite{Milo2002,Song2005}. For each network motif shown in Figure \ref{Motifs}, we compute the observability (controllability) indices for various measurement nodes, connection strengths, and driving inputs (dynamic regimes). Measurements of \(v_i\) for each motif were from each one of the nodes \(i=1, 2, \text{ or } 3\). Simulated network data were used to compute the observability (controllability) index for two cases: 1) where the system parameters for all 3 nodes and connections were identical, and 2) where the nodes had a heterogeneous (\(10\%\) variance) symmetry-breaking set of coupling parameters. To create simulated data, the full six-dimensional FN network equations were integrated from the same initial conditions with the same driving inputs for each node via a Runga-Kutta \(4^{th}\) order (RK4) method with time step \(\Delta t=0.04 \) for 12000 time steps (with the initial transient discarded) in MATLAB for each test case: 1) limit cycle and 2) chaotic dynamical regimes, with a) identical and b) heterogeneous coupling (the nodal parameters remain identical throughout). Convergence of solutions was achieved when \(\Delta t\) was decreased to \(0.004 \). Data were then imported into \textit{Mathematica} and inserted into symbolic observability and controllability matrices (computed for each node), which then were numerically computed to obtain the observability (controllability) indices for each coupling strength. The indices were then averaged over the integration paths starting from random initial conditions. These calculations are summarized in Figures \ref{M1Chaos} - \ref{M7Chaos}, \ref{CtrlbConsNS} and \ref{ObsLCNS} for observability and controllability, in the chaotic, pulsed limit cycle, and constant input limit cycle dynamical regimes.

\section{Results\label{results}}

\subsection{Motifs with Symmetry}
\begin{figure}[!h]
\begin{overpic}[height=3.3in,unit=1mm]{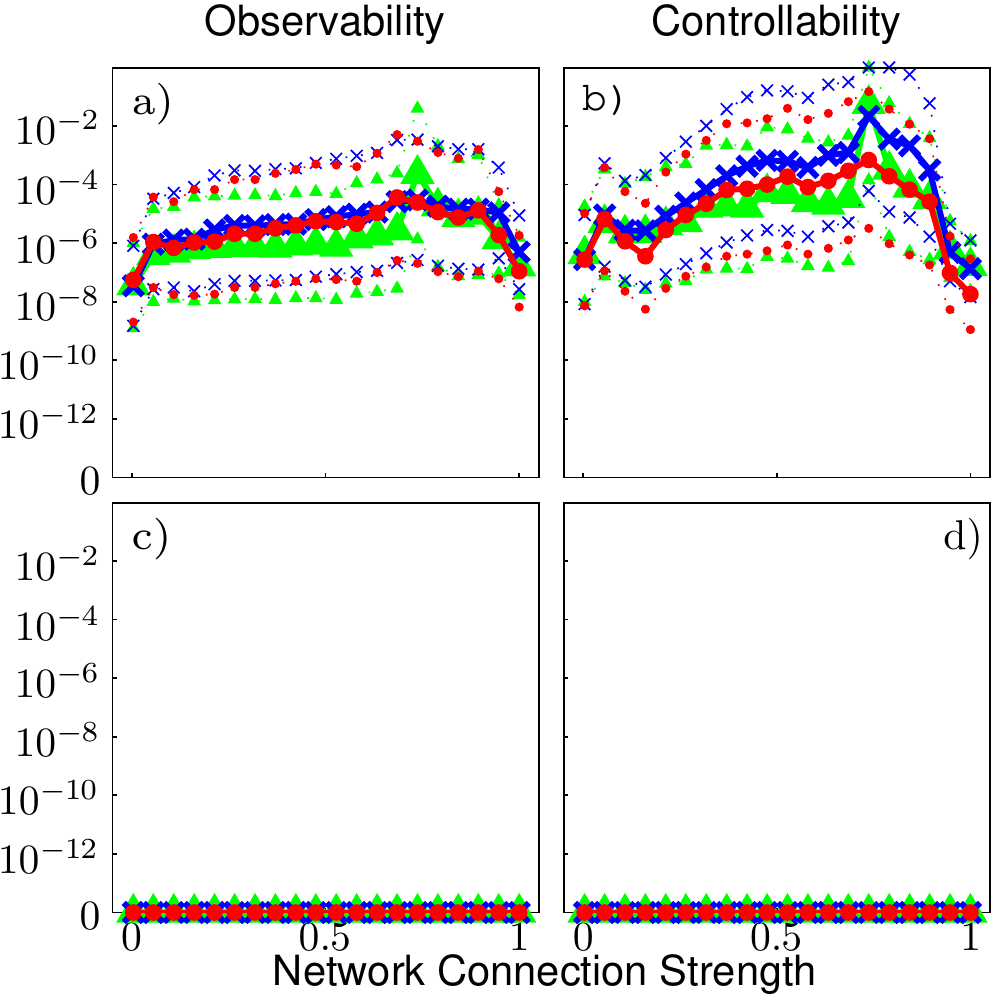}
	\put(38,33){\includegraphics[height=0.75in]{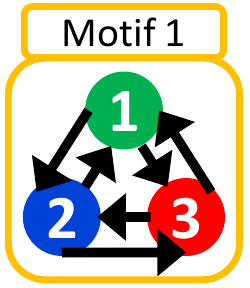}}
\end{overpic}
\caption{\label{M1Chaos}Calculation of observability (a and c) and controllability (b and d) indices for motif 1 for a chaotic dynamical regime, as measured from each node (green \(\triangle\) =1, blue \(\times\) =2, red \(\bullet\) =3). The thick lines and symbols mark the mean values of each distribution of indices for each coupling strength, while the smaller symbols and dotted lines represent the \(\pm 1\) standard deviation confidence intervals. Plots in the top row represent the results computed with symmetry breaking heterogeneous couplings while plots in the bottom row are those with identical coupling strengths. 
}
\end{figure}

\begin{figure}[!h]
\begin{overpic}[height=3.3in,unit=1mm]{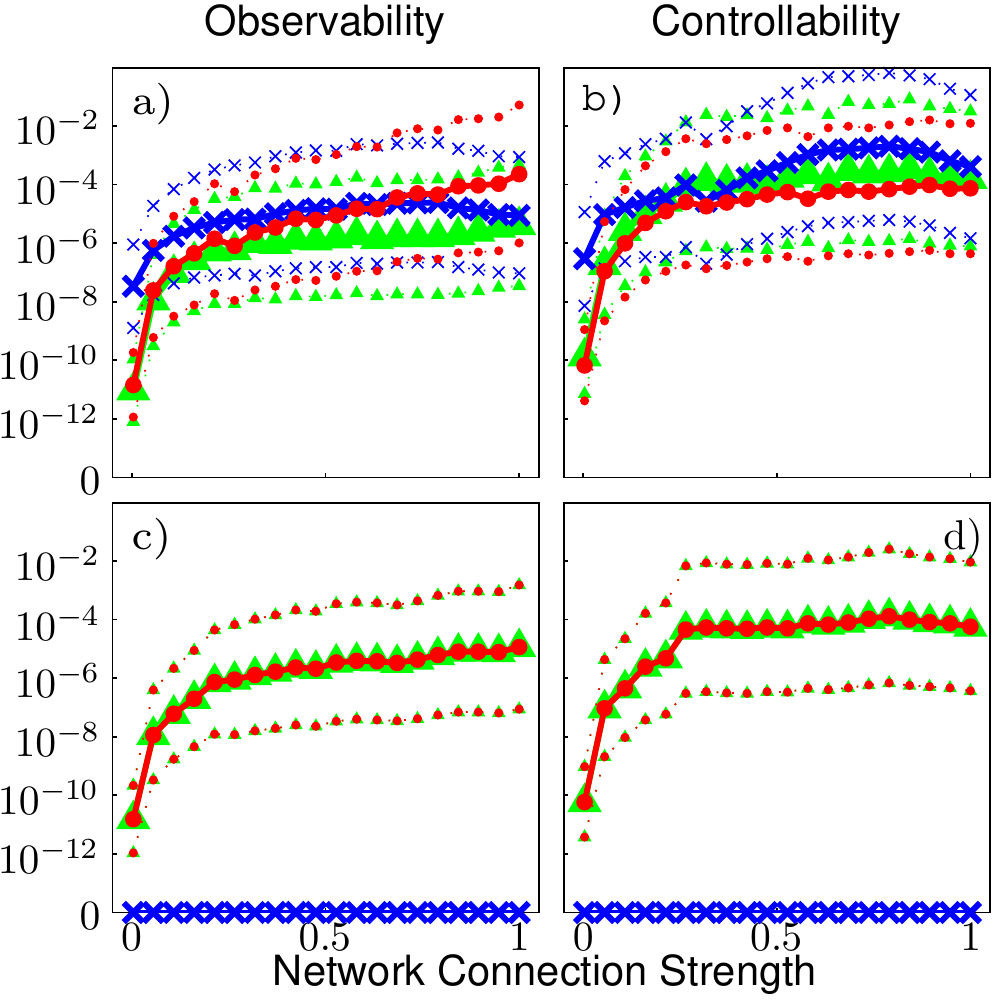}
	\put(38,33){\includegraphics[height=0.75in]{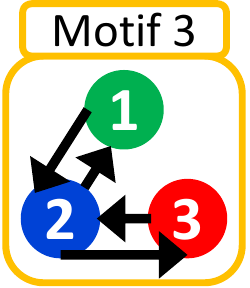}}
\end{overpic}
\caption{\label{M3Chaos}Same as Figure \ref{M1Chaos}, except calculations are for motif 3. The calculations show that the reflection symmetry in the network topology causes zero observability and controllability for the symmetric case of observing or controlling from node 2 with identical coupling strengths (c and d).}
\end{figure}

\begin{figure}[!h]
\begin{overpic}[height=3.3in,unit=1mm]{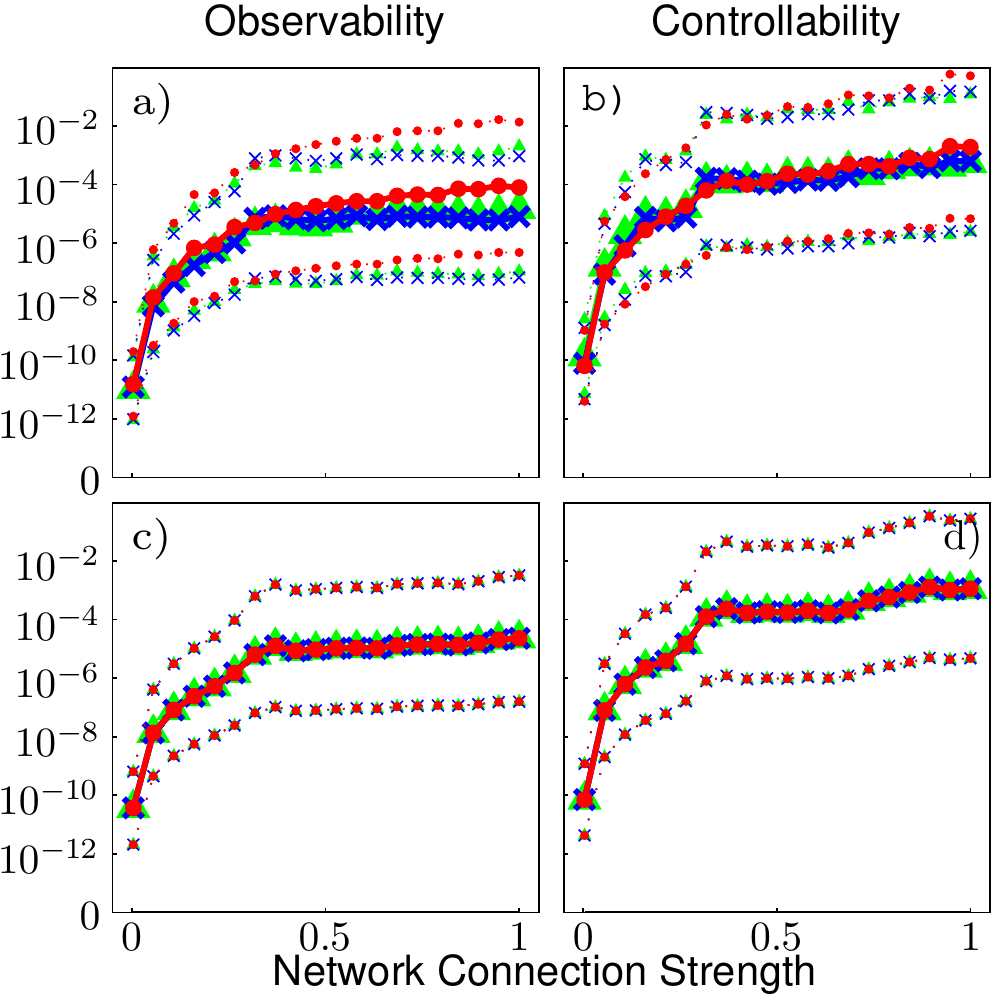}
	\put(38,33){\includegraphics[height=0.75in]{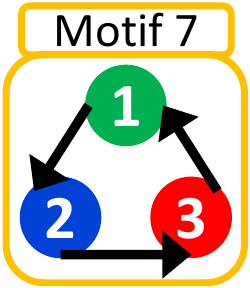}}
\end{overpic}
\caption{\label{M7Chaos}Same as Figure \ref{M1Chaos}, except calculations are for motif 7. The calculations show that the particular rotational symmetry in the network topology has no ill effect on observability and controllability for the symmetric case of identical coupling strengths (c and d) as compared to the broken symmetry in a and b.}
\end{figure}

For motif 1, the data show that a system with full \(\mathbf{S_3}\) symmetry (due to the connection topology and identical nodal and coupling parameters) generates zero observability (controllability) over the entire range of coupling strengths (Figure \ref{M1Chaos}c and \ref{M1Chaos}d). Similarly, no observability (controllability) is seen from node 2 in motif 3 which has a reflection  \(\mathbf{S_2}\) symmetry across the plane through node 2 (Figure \ref{M3Chaos}c and \ref{M3Chaos}d). Interestingly, the cyclic symmetry of motif 7 does not cause loss of observability (controllability) as shown in Figure \ref{M7Chaos}; motif 7 has rotational \(\mathbf{C_3}\) symmetry and valance 1 connectivity (1 input, 1 output). In motifs 1 and 3 the effect of the symmetry is partially broken by introducing a variation in the coupling terms, and the results show non-zero observability (controllability) indices in the plots for such heterogeneous coupling (plots a and b in Figures \ref{M1Chaos} and \ref{M3Chaos}) with a dependence on the coupling strength. 
 
Of particular interest is the substantial loss of observability (controllability) as the coupling strengths increase to critical levels for systems containing latent structural symmetries in the presence of heterogeneity (motifs 1 and 3, plots a and b in Figures \ref{M1Chaos} and \ref{M3Chaos}). That is, increasing the coupling strengths when recording (stimulating) from any node in motif 1 or node 2 in motif 3, degrades observability (controllability) as coupling strength increases. A study of the 3D phase plots of the FN voltage variable in motif 1 (as a function of coupling strength for chaotic dynamics) reveals a blowout bifurcation \cite{Ott1994} at lower values of coupling strengths (Figure \ref{3DChaos}), and at higher levels, generalized synchrony \cite{Schiff1996} and increased observability (controllability), and finally the subsequent decrease in observability (controllability) at the highest levels of coupling strength (motif 1 as observed (controlled) from any node in Figure \ref{M1Chaos}). This is demonstrated in motif 1 (Figure \ref{3DChaos}), where a bifurcation in the dynamics causes the wandering trajectories at weak coupling strengths to collapse onto the limit cycle attractor at stronger coupling strengths, and at the strongest coupling the dynamics reveal a reverse Hopf bifurcation from limit cycle back into a stable equilibrium. 

Although motif 7 contains symmetry, the observability and controllability measures appeared unaffected by the presence of this symmetry; further insight into why this happens in such networks requires group representation theory and is presented in section \ref{SymGroupTheory}.

\begin{figure*}
\includegraphics[width=7in, clip=true, trim = 0 0 0 0]{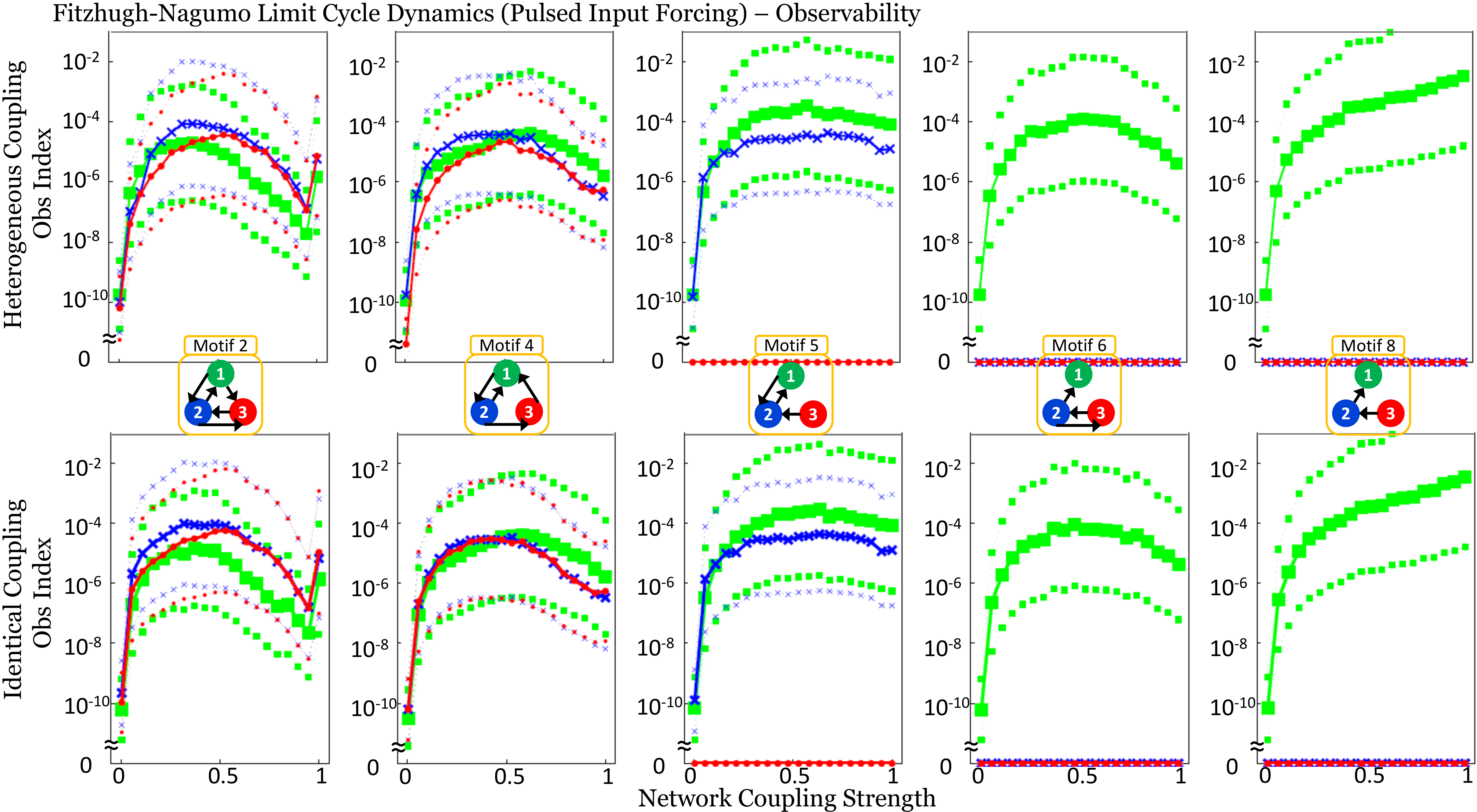}
\caption{\label{ObsLCNS}Calculation of observability indices for each of the FN network motifs with no underlying group symmetries for a pulsed input limit cycle dynamical regime, as measured from each node (green \(\Box\) =1, blue \(\times\) =2, red \(\bullet\) =3). The thick lines and symbols mark the mean values while the smaller symbols and dotted lines represent the \(\pm\)1 standard deviation confidence intervals. Plots in the top row are computed with heterogeneous couplings while identical coupling strengths are in the bottom row. The calculations show the effect of network coupling strength on observability; motifs 5, 6 and 8 show no observability from node 3 in motif 5, and nodes 2 and 3 in motifs 6 and 8 due to structural isolation.\\}
\includegraphics[width=7in, clip=true, trim = 0 0 0 -20]{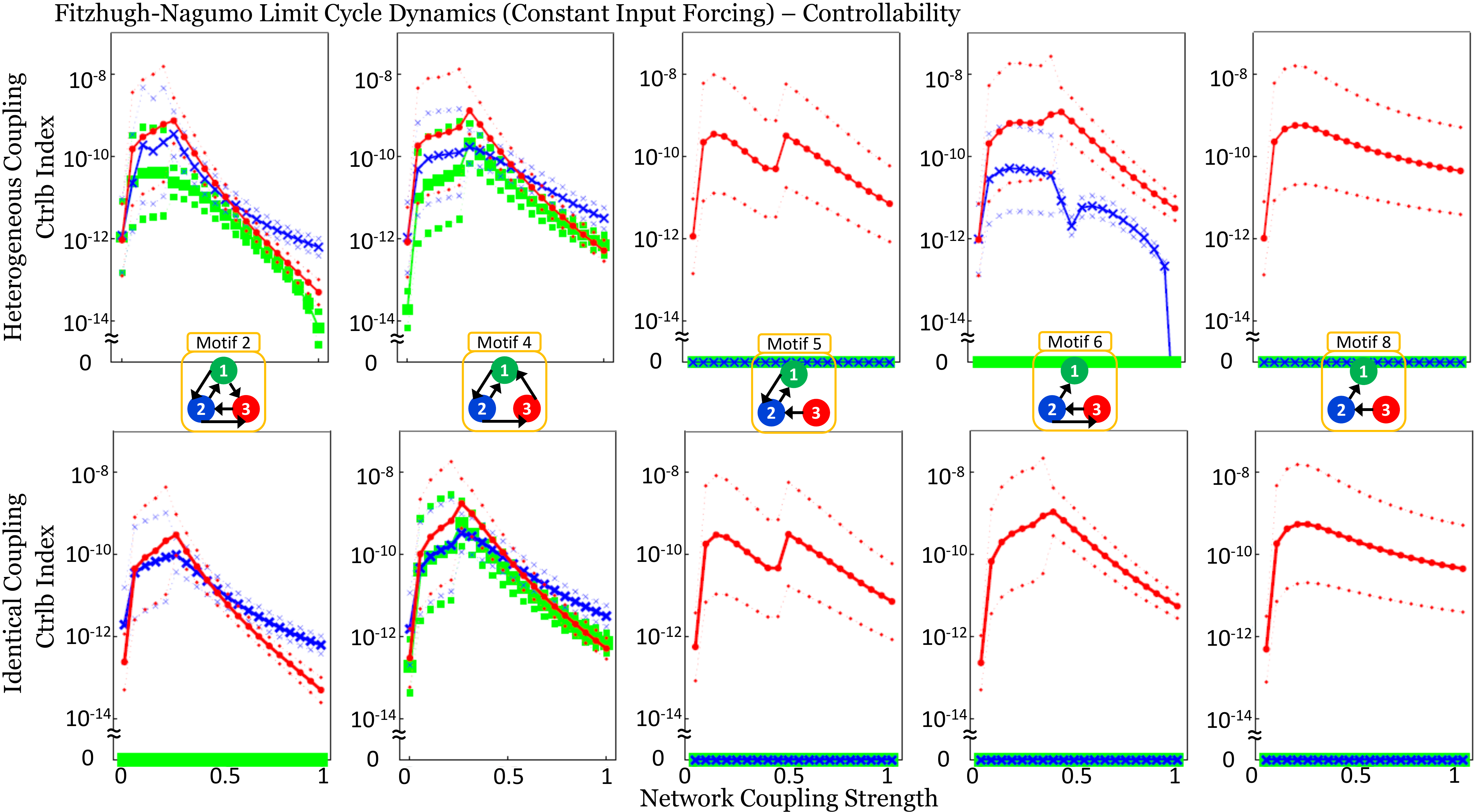}
\caption{\label{CtrlbConsNS}Calculation of controllability indices for each of the FN network motifs with no underlying group symmetries for a limit cycle dynamical regime with constant input current \(I=-0.45\), all other details are the same as in Figure \ref{ObsLCNS}. In particular, notice that local input-output symmetries cause zero controllability when controlling motif 2 from node 1 or motif 6 from node 2.}
\end{figure*}

\begin{turnpage}
\begin{figure*}
\includegraphics[width=9in, clip=true, trim = 10 10 0 5]{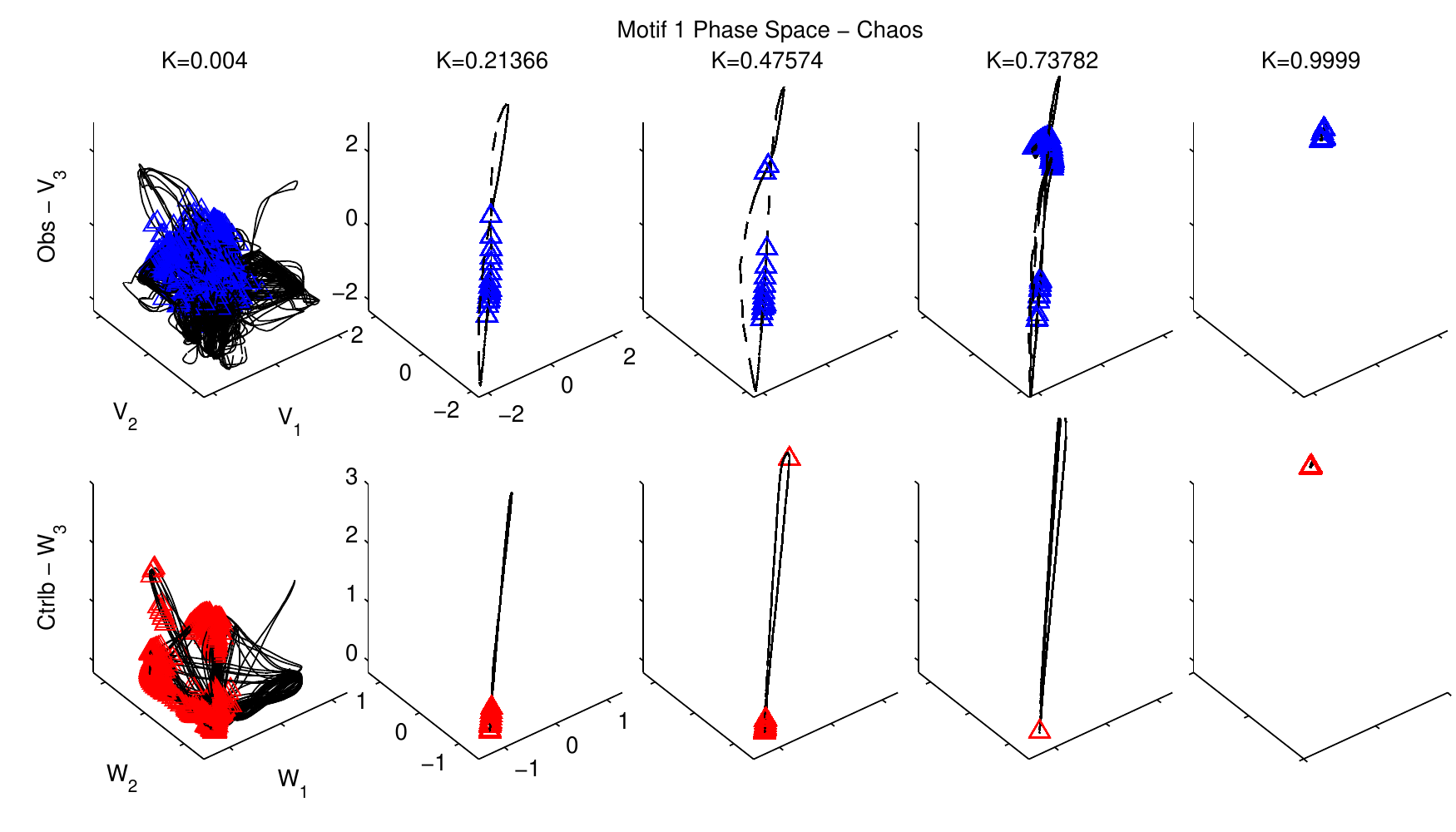}
\caption{\label{3DChaos}The 3-dimensional phase space for \(v\) and \(w\), showing trajectories in motif 1 as measured from node 1 for a range of connection strengths (weak to strong heterogeneous coupling K, from left to right respectively). In the first row, blue triangles mark locations in phase space where observability is higher than the mean for the trajectory, while the second row contains a phase space trajectory for \(w\) and red triangles mark the higher than average controllability. The broken symmetry of the heterogeneous network has trajectories that visit locations in the phase space that vary widely in observability and controllability with a log-normal distribution.}
\end{figure*}
\end{turnpage}

\subsection{Motifs without Symmetry}
Local output symmetries occur in motifs 2, and 6 when controlling from the first and second node respectively (green and blue traces in Figure \ref{CtrlbConsNS}), which is remedied by the disambiguating effect of parameter variation. Additionally, as in the motifs with symmetry, the broken local symmetries lose controllability as coupling strength further increases evident in motifs 2 and 6 in Figure \ref{CtrlbConsNS}. In the cases where the indices are zero without symmetries (motifs 5, 6, and 8 in Figures \ref{ObsLCNS} and \ref{CtrlbConsNS}), the motif must contain one or more structurally isolated nodes and hence are not structurally controllable or observable. From the viewpoint of observability this means that information from the isolated node(s) cannot reach the measured node as the two are not connected in that direction \cite{Rech1990,Joly2012}; for controllability, this means that the isolated node(s) is not reached by the controlled node due to the two not being connected in that direction \cite{Lin1974}. This structural nodal isolation is exemplified in motif 8 (in Figures \ref{ObsLCNS} and \ref{CtrlbConsNS}), where the network is only observable from node 1, and only controllable from node 3.

Additionally, the plots in Figures \ref{ObsLCNS} and \ref{CtrlbConsNS} show counter-intuitively that as coupling strength increases the observability (controllability) indices can increase to an optimal value, and then begin to decrease as coupling strength increases past this critical coupling value. 
\clearpage
\begin{figure}
\includegraphics[width=1.1in, clip=true, trim = 0 0 0 0]{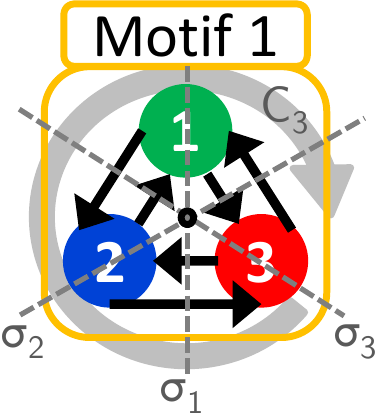}
\caption{\label{symFig}Graphic illustration of symmetry axes \(\sigma_n\) with \(n=1,2,3\) and the cyclic rotation symmetry \(C_3\) about an axis perpendicular to the plane of the page.}
\end{figure}

\section{Symmetric Network Observability and Controllability via Group Representation Theory\label{SymGroupTheory}}
For linear time-varying systems, Rubin \& Meadows \cite{Rubin1972} used the theory of group representations \cite{Burnside1955,Wigner1959,Hamermesh1962,Tinkham1964} to show how a (circuit) network containing group symmetries would be non-controllable or non-observable due to symmetries (termed NCS or NOS respectively). The analysis involves first determining the irreducible representations of the symmetry group of the system equations, then constructing an orthogonal basis (called a symmetry basis) from the irreducible representations which transforms the system matrix \(A(t)\) into block diagonal form (also called modal form). Inspection of the fully transformed system from (\ref{LTIsys}) will reveal if the NCS or NOS property is present via zeros in a critical location of decoupled block-diagonal decomposition \((\hat{A},\hat{B},\hat{C})\) i.e. the form
\begin{equation}
\label{Xsys}
\begin{gathered}
	\frac{d}{dt}
	\begin{bmatrix}
		Z_1 \\ Z_2
	\end{bmatrix}
	=\underbrace{
	\begin{bmatrix}
		A_1 & 0 \\ 0 & A_2
	\end{bmatrix}}_\text{\(\hat{A}\)}
	\begin{bmatrix}
		Z_1 \\ Z_2
	\end{bmatrix}
	+\underbrace{
	\begin{bmatrix}
		B_1 \\ 0
	\end{bmatrix}}_\text{\(\hat{B}\)}
	u(t) \\
	y(t)=\underbrace{
	\begin{bmatrix}
		C_1 & 0
	\end{bmatrix}}_\text{\(\hat{C}\)}
	\begin{bmatrix}
		Z_1 \\ Z_2
	\end{bmatrix},
\end{gathered}
\end{equation}
where the transformed system (\ref{Xsys}) in partitioned form above is non-controllable and non-observable (not completely controllable or observable). This can be seen by inspection, as the zeros present in the partitioned measurement and control functions \(\hat{C}\) and \(\hat{B}\) leave the transformed system unable to measure or control the mode associated with \(Z_2\) as neither \(u(t)\) or \(Z_1\) is present in the equation for \(Z_2\) and \(Z_2\) does not appear in the output. In the next section we summarize the minimum background components of groups and representations (without proofs) in order to further gain insight into how symmetry effects the controllability and observability of our networks. 

\subsection{Symmetric Groups and Representations}
A symmetry operation on a network is a permutation (in this case nodes) that results in exactly the same configuration as before the transformation was applied. The symmetric group \(\mathbf{S_n}\) consists of all permutations on \(n\) symbols - called the order of the group \(g\). The shorthand method of denoting a permutation operation \(R\) of nodes in a network will be written \((123)\), where node 1 is replaced by node 2 and node 2 by node 3. This is called a cycle of the permutation \cite{Burnside1955}, and with it we can define all of the permutations of \(\mathbf{S_n}\). Three of the network motifs studied here contain topological symmetries (Figures \ref{M1Chaos}, \ref{M3Chaos} and \ref{M7Chaos}); motif 1 has \(\mathbf{S_3}\) symmetry, motif 3 has \(\mathbf{S_2}\) symmetry and motif 7 contains \(\mathbf{C_3}\) symmetry\footnote{See \cite{Tinkham1964} for a rigorous classification of various forms of symmetry.}, and each of these groups comprise the following sets of permutation operations \(R\)
\begin{equation}
\label{R}
	\begin{aligned}
		R:\mathbf{S_3} = \{&E,\sigma_1,\sigma_2,\sigma_3,C_3,C_3^2 \} \\
		=\{&E=(1)(2)(3) \\
		&\sigma_1=(23), \sigma_2=(13), \sigma_3=(12) \\
		&C_3=(132), C_3^2=(123) \},
	\end{aligned}
\end{equation}
where \(E\) is the identity operation, \(\sigma_n\) is a reflection across the \(n^{\text{th}}\) axis in Figure \ref{symFig}, and \(C_3\) and \(C_3^2\) are two cyclic rotations where \(C_n\) denotes a rotation of the system by \(2\pi/n\) radians where the system remains invariant after rotation \cite{Tinkham1964}. \(\mathbf{S_2}\) and \(\mathbf{C_3}\) symmetry in motifs 3 and 7 respectively are subgroups of \(\mathbf{S_3}\)
\begin{equation}
\label{R2}
	\begin{aligned}
		\mathbf{S_2} &= \{E,\sigma_2 \} \\
		\mathbf{C_3} &= \{E,C_3,C_3^2 \}
	\end{aligned}
\end{equation}
The permutation operations \(R\) in these symmetric groups can also be represented by monomial matrices\footnote{A monomial matrix has only one non-zero entry per row and column. In this case permutation operations limit those values to either +1 or -1. } \(D(R)\):
\begin{equation}
\label{DR3}
\begin{alignedat}{6}
	&\begin{bmatrix}
		1 \!\!& 0 \!\!& 0 \\
		0 \!\!& 1 \!\!& 0 \\
		0 \!\!& 0 \!\!& 1 
	\end{bmatrix}\!\! &
	&\begin{bmatrix}
		1 \!\!& 0 \!\!& 0 \\
		0 \!\!& 0 \!\!& 1 \\
		0 \!\!& 1 \!\!& 0 
	\end{bmatrix}\!\! &
	&\begin{bmatrix}
		0 \!\!& 0 \!\!& 1 \\
		0 \!\!& 1 \!\!& 0 \\
		1 \!\!& 0 \!\!& 0 
	\end{bmatrix}\!\! &
	&\begin{bmatrix}
		0 \!\!& 1 \!\!& 0 \\
		1 \!\!& 0 \!\!& 0 \\
		0 \!\!& 0 \!\!& 1 
	\end{bmatrix}\!\! &
	&\begin{bmatrix}
		0 \!\!& 1 \!\!& 0 \\
		0 \!\!& 0 \!\!& 1 \\
		1 \!\!& 0 \!\!& 0 
	\end{bmatrix}\!\! &
	&\begin{bmatrix}
		0 \!\!& 0 \!\!& 1 \\
		1 \!\!& 0 \!\!& 0 \\
		0 \!\!& 1 \!\!& 0 
	\end{bmatrix} \\
	&\quad\;\; E & &\quad\;\; \sigma_1 & &\quad\;\; \sigma_2 & &\quad\;\; \sigma_3 & &\quad\;\; C_3 & &\quad\;\; C_3^2
\end{alignedat}
\end{equation}
where D(R) in (\ref{DR3}) is a 3-dimensional representation of \(\mathbf{S_3}\) group symmetry (for our 3 node motifs); a representation \(D(R)\) for \(\mathbf{S_2}\) and \(\mathbf{C_3}\) group symmetry are just the matrices above in (\ref{DR3}) corresponding to the sets of group elements given in (\ref{R2}).

A group of matrices \(D(\cdot)\) is said to form a representation of a group \(\mathbf{S_n}\) if a correspondence (denoted \(\sim\)) exists between the matrices and the group elements such that products correspond to products, i.e., if \(R_1\sim D(R_1)\) and \(R_2\sim D(R_2)\), then the composition \((R_1 R_2)\sim D(R_1) D(R_2)=D(R_1 R_2)\) (\textit{Definition 12} in \cite{Rubin1972}); this is known as a homomorphism of the group to be represented, and if the correspondence is one-to-one the representation is isomorphic and called a ``faithful" representation of the group.

\textit{Theorem 2} from \cite{Rubin1972} establishes the connection between group theory and the linear network system equations (\ref{LTIsys}), by demonstrating that the monomial representation \(D(R)\) of symmetry operations \(R\) is conjugate (commutes) with the network system matrix \(A\) in (\ref{LTIsys}):
\begin{equation}
	D^{-1}(R)A(t)D(R)=A(t), \quad \forall R \in \mathbf{S_n}
\end{equation}
where \(D(R)\) shows how the states of the system equations transform under the symmetry operation \(R\), and form a reducible representation \cite{Burnside1955,Kerns1951} of the symmetric group \(\mathbf{S_n}\). A representation is said to be reducible if it can be transformed into a block diagonal form via a similarity transformation \(\alpha\), and irreducible if it is already in diagonal form; a reducible representation \(D(R)\) that has been reduced to block diagonal form \(\hat{D}(R)\) will have \(k\) non-zero submatrices along the diagonal that define the irreducible representations \(D^{(p)}(R), p=1\hdots k \) of the group \(\mathbf{S_n}\) \cite{Rubin1972}

\begin{equation}
\label{aDa}
\arraycolsep=1.4pt\def\arraystretch{0.1}
	\begin{gathered}
		\alpha^{\dag}D(R)\alpha=\hat{D}(R), \quad \forall R \in \mathbf{S_n} \\
		\hat{D}(R)=	\begin{bmatrix} D_{l_1}^{(1)}\!\! &\!\! & \!\!\!\!\!0 \\ & \!\!\ddots & \!\! \\ \;0\!\!& \!\! & \!\!D_{l_k}^{(k)}\! \end{bmatrix},
	\end{gathered}
\end{equation}
where \(\dag\) represents the complex conjugate transpose of \(\alpha\), \(l_p\) is the dimension of \(D^{(p)}(R)\) and the number of irreducible representations \(k\) equals the number of classes the group elements \(R\) are partitioned into. This can be found by computing the trace of each representation in \(D(R), \: \forall R\) - called the character of the representation - and collecting those that have the same trace into separate classes \(\mathscr{C}_p, p=1\hdots k\), which define sets of conjugate elements \cite{Tinkham1964}. The character of \(D(R)\) is defined as
\begin{equation}
\label{char}
	\chi(R)=\text{Tr}(D(R)), \quad \forall R \in \mathbf{S_n}.
\end{equation}
The key to forming irreducible representations in (\ref{aDa}) is that the transform \(\alpha\) needs to reduce each representation matrix \(D(R)\) to diagonal form for every group element \(R\) in \(\mathbf{S_n}\).

In (\ref{aDa}) the dimension of each irreducible representation \(l_p\) can be found from the fact that the irreducible representations of the group form an orthogonal basis in the \(g\)-dimensional space of the group, and since there can be no more than \(g\) independent vectors in the orthogonal basis it can be shown \cite{Hamermesh1962} that
\begin{equation}
\label{dim}
	\sum_{p=1}^{k}l_p^2=g,
\end{equation}
where the sum is over the number of irreducible representations (or classes of conjugate group elements) \(k\). Some of the irreducible representations \(D^{(p)}(R)\) will appear in \(\hat{D}(R)\) more than once while others may not appear at all; the character of the representation completely determines this and the number of times, \(a_p\), that \(D^{(p)}(R)\) appears in \(\hat{D}(R)\) is defined in \cite{Tinkham1964} as
\begin{equation}
\label{ap}
	a_p=\frac{1}{g}\sum_{R}\chi^{(p)}(R)^*\chi(R),
\end{equation}
where \(\chi^{(p)}(R)\) is the trace of \(D^{(p)}(R)\), the asterisk denotes complex conjugate and \(\chi(R)\) is the trace of \(D(R)\). 

\subsection{Construction of the Similarity Transform\footnote{For purposes of clarity, we simplified the presentation of the computation of \(\alpha\) for our motifs where there is only one set of network nodes that can permuted amongst themselves. For the more general case where the group operations \(R\) are separated into subgroups corresponding to different sets of permutable network nodes (e.g. RLC networks, or different neuron types) see \cite{Rubin1972}.} \(\alpha\)\label{ConstrAlpha}}

We examine motif 3 in Figure \ref{M3Chaos} which has \(\mathbf{S_2}\) symmetry. Determined from (\ref{char}), there are 2 classes of group elements \(\mathscr{C}_1=\{E\} \text{ and }\mathscr{C}_2=\{\sigma_2\}\), and reduction of \(D(R)\) yields the two, 1-dimensional (\(l_1=l_2=1\) computed from (\ref{dim})) irreducible representations \(D^{(1)}(R) \text{ and }D^{(2)}(R)\) of \(\mathbf{S_2}\):
\begin{equation}
\label{IrrS2}
\renewcommand{\arraystretch}{1.5}
\setlength{\tabcolsep}{8pt}
\begin{tabular}{c | c c}
	R & E & $\sigma_2$ \\ \hline
	$D^{(1)}(R)$ & 1 & 1 \\
	$D^{(2)}(R)$ & 1 &-1
\end{tabular}
\end{equation}
where each entry in \(D^{(p)}\) corresponds to the elements of \(D(R)\) above in equation (\ref{DR3}), where \(R = \{E,\sigma_2 \} \) as in equation (\ref{R2}), and from equation (\ref{ap}), \(D^{(1)}(R)\) appears two times while \(D^{(2)}(R)\) appears once in \(D(R)\).

A procedure for transforming the reducible representation \(D(R)\) of a symmetry group \(\mathbf{S_n}\) to block diagonal form is presented in \cite{Kerns1951,Rubin1972}. A unitary transformation \(\alpha\) is constructed from the normalized linearly independent columns of the \(n \times n\) generating matrix \(G_{i}^{(p)}\)
\begin{equation}
\label{G}
		G_{i}^{(p)}=\sum_{R} D^{(p)}(R)_{i i}^{*}D(R),
\end{equation}
where \(D^{(p)}(R)_{i i}\) is the \((i,i)^{th}\) diagonal entry of a \(l_p\)-dimensional irreducible representation \(p\) (hence \(i=1\hdots l_p\)) of the symmetry group \(\mathbf{S_n}\) and the asterisk denotes complex conjugate. Each matrix \(G_i^{(p)}\) will contribute \(a_p\) linearly independent columns from (\ref{ap}) to form the coordinate transformation matrix \(\alpha\). Using equations (\ref{IrrS2}) and (\ref{G}) and iterating through all \(l_p\) rows of each of the \(k\) irreducible representations in (\ref{aDa}), we construct \(\alpha\) for motif 3
\begin{equation}
\renewcommand{\arraystretch}{1.0}
\begin{aligned}
	G_{1}^{(1)}&=\sum_{R\in S_2} D^{(1)}(R)_{1 1}^{*}D(R) \\
	&=1\begin{bmatrix}
		1 \!\!& 0 \!\!& 0 \\
		0 \!\!& 1 \!\!& 0 \\
		0 \!\!& 0 \!\!& 1 
	\end{bmatrix}\!\! +
	1\begin{bmatrix}
		0 \!\!& 0 \!\!& 1 \\
		0 \!\!& 1 \!\!& 0 \\
		1 \!\!& 0 \!\!& 0 
	\end{bmatrix} 
	=\begin{bmatrix}
		1 \!\!& 0 \!\!& 1 \\
		0 \!\!& 2 \!\!& 0 \\
		1 \!\!& 0 \!\!& 1 
	\end{bmatrix},
\end{aligned}
\end{equation}
where each linearly independent column of \(G\) is a column of \(\alpha\). After normalizing we have
\begin{equation}
\renewcommand{\arraystretch}{1.0}
	\begin{bmatrix}
		1 \\ 0 \\ 1
	\end{bmatrix},
	\begin{bmatrix}
		0 \\ 2 \\ 0
	\end{bmatrix}
	\xrightarrow[\text{normalize}]{ }
	\begin{bmatrix}
		\frac{1}{\sqrt{2}} \\ 0 \\ \frac{1}{\sqrt{2}}
	\end{bmatrix},
	\begin{bmatrix}
		0 \\ 1 \\ 0
	\end{bmatrix}
	=\begin{bmatrix}
		\alpha_{11} & \alpha_{21} \\ \alpha_{12} & \alpha_{22} \\ \alpha_{13} & \alpha_{23}
	\end{bmatrix},
\end{equation}
which defines the first and second columns of \(\alpha\). Continuing, we have
\begin{equation}
\renewcommand{\arraystretch}{1.0}
\begin{aligned}
	G_{1}^{(2)}&=\sum_{R\in S_2} D^{(2)}(R)_{1 1}^{*}D(R) \\
	&=1\begin{bmatrix}
		1 \!\!& 0 \!\!& 0 \\
		0 \!\!& 1 \!\!& 0 \\
		0 \!\!& 0 \!\!& 1 
	\end{bmatrix}\!\! -
	1\begin{bmatrix}
		0 \!\!& 0 \!\!& 1 \\
		0 \!\!& 1 \!\!& 0 \\
		1 \!\!& 0 \!\!& 0 
	\end{bmatrix} 
	=\begin{bmatrix}
		1 \!\!& 0 \!\!& -1 \\
		0 \!\!& 0 \!\!& 0 \\
		-1 \!\!& 0 \!\!& 1 
	\end{bmatrix},
\end{aligned}
\end{equation}
which yields the final column of \(\alpha\) (after normalization)
\begin{equation}
\renewcommand{\arraystretch}{1.0}
	\begin{bmatrix}
		1 \\ 0 \\ -1
	\end{bmatrix}
	\xrightarrow[\text{normalize}]{ }
	\begin{bmatrix}
		\frac{1}{\sqrt{2}} \\ 0 \\ -\frac{1}{\sqrt{2}}
	\end{bmatrix}
	=\begin{bmatrix}
		\alpha_{31} \\ \alpha_{32} \\ \alpha_{33}
	\end{bmatrix}.
\end{equation}
Now the coordinate transformation matrix \(\alpha\) is
\begin{equation}
\renewcommand{\arraystretch}{1.0}
	\alpha=
	\begin{bmatrix}
		\frac{1}{\sqrt{2}} & 0 & \frac{1}{\sqrt{2}} \\
		0 & 1 & 0 \\
		\frac{1}{\sqrt{2}} & 0 & -\frac{1}{\sqrt{2}}
	\end{bmatrix}.
\end{equation}
Motif 3 in Figure \ref{M3Chaos} has connection matrix \(A_3\)
\begin{equation}
\label{AM3}
\renewcommand{\arraystretch}{1.0}
	A_3=
	\begin{bmatrix}
		0 & 1 & 0 \\
		1 & 0 & 1 \\
		0 & 1 & 0
	\end{bmatrix}.
\end{equation}
To control from node 1,2 and 3 respectively, the \(B\) matrix takes the form
\begin{equation}
\label{B123}
\renewcommand{\arraystretch}{1.0}
	B_{1,2,3}=
	\begin{bmatrix}
		 1 \\ 0 \\ 0
	\end{bmatrix},
	\begin{bmatrix}
		 0 \\ 1 \\ 0
	\end{bmatrix},
	\begin{bmatrix}
		 0 \\ 0 \\ 1
	\end{bmatrix},
\end{equation}
and to observe from node 1,2 and 3 respectively, the \(C\) matrix takes the form
\begin{equation}
\label{C123}
\renewcommand{\arraystretch}{1.0}
	C_{1,2,3}=
	\begin{bmatrix}
		 1 & 0 & 0
	\end{bmatrix},
	\begin{bmatrix}
		 0 & 1 & 0
	\end{bmatrix},
	\begin{bmatrix}
		 0 & 0 & 1
	\end{bmatrix}.
\end{equation}
The block diagonalized system  \((\hat{A}_3,\hat{B},\hat{C})\) is formed with the substitution \(Z=\alpha^{\dag}x\), and (\(A_3,B,C\)) in (\ref{AM3}) to (\ref{C123}) becomes
\begin{equation}
\renewcommand{\arraystretch}{1.0}
\label{Xsys2}
\begin{gathered}
	\hat{A}_3:\alpha^{\dag}A_3\alpha =
	\begin{bmatrix}
		0 & \sqrt{2} & 0 \\
		\sqrt{2} & 0 & 0 \\
		0 & 0 & 0
	\end{bmatrix} \\
	\hat{B}:\alpha^{\dag}B_{1,2,3}=
	\begin{bmatrix}
		 \frac{1}{\sqrt{2}} \\ 0 \\ \frac{1}{\sqrt{2}}
	\end{bmatrix},
	\begin{bmatrix}
		 0 \\ 1 \\ 0
	\end{bmatrix},
	\begin{bmatrix}
		 \frac{1}{\sqrt{2}} \\ 0 \\ \frac{-1}{\sqrt{2}}
	\end{bmatrix} \\
	\hat{C}:C_{1,2,3}\alpha=
	\begin{bmatrix}
		 \frac{1}{\sqrt{2}} & 0 & \frac{1}{\sqrt{2}}
	\end{bmatrix},
	\begin{bmatrix}
		 0 & 1 & 0
	\end{bmatrix},
	\begin{bmatrix}
		 \frac{1}{\sqrt{2}} & 0 & \frac{-1}{\sqrt{2}}
	\end{bmatrix}
\end{gathered}
\end{equation}
By inspection of the transformed system (\ref{Xsys2}) it becomes clear that motif 3 is non-controllable and non-observable from node 2 due to symmetry alone (NCS and NOS), i.e. the transformed system in modal coordinates
\begin{equation}
\renewcommand{\arraystretch}{1.0}
\label{Xsys3}
\begin{gathered}
	\frac{d}{dt}
	\begin{bmatrix}
		Z_1 \\ Z_2 \\Z_3
	\end{bmatrix}
	=
	\begin{bmatrix}
		0 & \sqrt{2} & 0 \\
		\sqrt{2} & 0 & 0 \\
		0 & 0 & 0
	\end{bmatrix}
	\begin{bmatrix}
		Z_1 \\ Z_2 \\ Z_3
	\end{bmatrix}
	+
	\begin{bmatrix}
		0 \\ 1 \\ 0
	\end{bmatrix}
	u(t) \\
	y(t)=
	\begin{bmatrix}
		 0 & 1 & 0
	\end{bmatrix}
	\begin{bmatrix}
		Z_1 \\ Z_2 \\ Z_3
	\end{bmatrix},
\end{gathered}
\end{equation}
is NCS and NOS as the mode associated with \(Z_3\) cannot be reached by the input \(\hat{B}_2\) nor can its measurement be inferred from the output \(\hat{C}_2\) as in (\ref{Xsys}). 

The procedure to reduce motif 1 is accomplished in similar fashion\footnote{Full computation of \(\alpha\) is detailed in the appendix.} and the connection matrix \(A_1\) and its reduced form \(\hat{A_1}\) is:
\begin{equation}
\label{A1}
\renewcommand{\arraystretch}{1.0}
	A_1=
	\begin{bmatrix}
		0 & 1 & 1 \\
		1 & 0 & 1 \\
		1 & 1 & 0
	\end{bmatrix}, \;\;
	\hat{A}_1=
	\begin{bmatrix}
		2 & 0 & 0 \\
		0 & -1 & 0 \\
		0 & 0 & -1
	\end{bmatrix}
\end{equation}
while the transformed \(B \text{ and } C\) matrices in (\ref{B123}) and (\ref{C123}) are:
\begin{equation}
\renewcommand{\arraystretch}{1.0}
\label{XsysA1}
\hat{B}_{1,2,3}=
	\begin{bmatrix}
		 \frac{1}{\sqrt{3}} \\ \sqrt{\frac{2}{3}} \\ 0
	\end{bmatrix},
	\begin{bmatrix}
		 \frac{1}{\sqrt{3}} \\ \frac{-1}{\sqrt{6}} \\ \frac{1}{\sqrt{2}}
	\end{bmatrix},
	\begin{bmatrix}
		 \frac{1}{\sqrt{3}} \\ \frac{-1}{\sqrt{6}} \\ \frac{-1}{\sqrt{2}}
	\end{bmatrix}, 
	\hat{C}_{123}= \hat{B}_{1,2,3}^{T}
\end{equation}
At first glance it appears that motif 1 is NCS and NOS for measurement and control from node 1 only, and fully controllable and observable from node 2 and 3, however there is a subtle nuance to the controllability and observability of the diagonal form used in \cite{Rubin1972} and consolidated in (\ref{Xsys}) to show non-controllability and non-observability by inspection. 

It is well known that every non-singular \(n \text{ x } n\) matrix has \(n\) eigenvalues \(\lambda_n\) and \(n\) linearly independent eigenvectors, and that a matrix with repeated eigenvalues of algebraic multiplicity \(m_i\) will have a degeneracy \(1\leq q_i \leq m_i\) associated with the number of linearly independent eigenvectors for repeated eigenvalue \(\lambda_i\). This degeneracy \(q_i\) is also called the geometric multiplicity of \(\lambda_i\), and is equal to the dimension of the null space of \(A-I\lambda_i\) \cite{Brogan1974}. When utilizing similarity transforms to reduce a matrix to diagonal (modal) form this degeneracy in the eigenvectors (brought about by repeated eigenvalues) results in a transformed matrix that is almost diagonal, called the Jordan form matrix. The Jordan form is comprised of submatrices of dimension \(m_i\) - called Jordan blocks -  that have ones on the super-diagonal of each Jordan block \(J_i\) associated with the generalized eigenvectors of a repeated eigenvalue \(\lambda_i\). The diagonal form in (\ref{Xsys}) is a special case of Jordan form where the matrices on the diagonal are Jordan blocks of dimension one. This is known as the fully degenerate case with \(q_i=m_i\), and the Jordan form will have \(m_i\) separate \(1 \text{ x } 1\) Jordan blocks associated with each eigenvalue \(\lambda_i\).

The observability and controllability of systems in Jordan form hinges on where the zeros appear in the partitioned \(C_i\) and \(B_i\) matrices, where subscript \(i\) indicates a partition associated with a particular Jordan block \(J_i\). Given in \cite{Brogan1974,Bay1999} the conditions for controllability and observability of a system in Jordan form are:
\begin{enumerate}
	\item The first columns of \(C_i\) or the last rows of \(B_i\) must form a linearly independent set of vectors \(\{c_{1 1}\hdots c_{1 q_i}\} \text{ or } \{b_{1 e}\hdots b_{q_i e}\}\) (subscript \(e\) indicates the last row) corresponding to the \(q_i\) Jordan blocks \(J_1^{\lambda_i} \hdots J_{q_i}^{\lambda_i}\) for repeated eigenvalue \(\lambda_i\)
	\item \(c_{1 p} \neq 0\) or \(b_{p e} \neq 0\) when there is only one Jordan block \(J_p^{\lambda_i}\) associated with eigenvalue \(\lambda_i\)
	\item For single output and single input systems, the partitions of \(C_i\) and \(B_i\) are scalars - which are never linearly independent - thus each repeated eigenvalue must only have one Jordan block \(J_i^{\lambda_i}\) associated with it for observability or controllability respectively.
\end{enumerate}
From these criteria, we can now see that the transformed system for motif 1 in (\ref{A1}) contains three \(1 \text{ x } 1\) Jordan blocks, two of which are associated with the repeated eigenvalue \(\lambda_2=-1\), which violates condition 3); thus we conclude it is NCS and NOS.

\subsection{Motif 7 and Networks Containing Only Rotation Groups}

In \cite{Rubin1972}, it was shown how the \(r\)th component of \(\alpha\) vanishes according to the matrices \(D^{(p)}(R_r^r)\), where \(R_r^r\) represents a subgroup of the group operations (\(R\)) that transform the \(r\)th state variable into itself. Subsequently, two theorems were proven that make use of this fact to simplify the analysis of networks that have a single input or output coupled only to the \(r\)th state variable, which is precisely parallel to our analysis in section \ref{results}. A paraphrasing of \textit{Theorem 6 } and \textit{12} from \cite{Rubin1972} for controllability and observability states that such a single input or output network is NCS or NOS if and only if there is an irreducible representation \(D^{(p)}(R)\) that appears in \(D(R)\) and
\begin{equation}
\label{Thm6}
	\sum_{R_r^r}s_r^r D^{(p)}(R_r^r)_{ii}^*=0
\end{equation}
for some value of \(i\), where \(s_r^r\) is \(+1 \text{ or }-1\) as \(R_r^r\) transforms state variable \(x_r\) into itself with a plus or minus sign\footnote{in our motifs \(D(R)\) is a permutation representation, thus \(s_r^r=+1\).}. For this theorem to hold, the equality in (\ref{Thm6}) must be checked for all possible \(p\) for \(D^{(p)}(R)\) that appear in \(D(R)\) via (\ref{ap}). 

Applying (\ref{Thm6}) to motif 7, the irreducible representations for \(\mathbf{C_3}\) symmetry are:
\begin{equation}
\label{IrrC3}
\renewcommand{\arraystretch}{1.5}
\setlength{\tabcolsep}{8pt}
\begin{tabular}{c | c c c}
	R & E & $C_3$ & $C_3^2$ \\  \hline
	$D^{(1)}(R)$ & 1 & 1 & 1\\
	$D^{(2)}(R)$ & 1 & $\omega$ & $\omega^2$\\
	$D^{(3)}(R)$ & 1 & $\omega^2$ & $\omega$
\end{tabular}
\end{equation}
where \(\omega=e^{\frac{2\pi i}{3}}\). From the subset (\ref{R2}) of (\ref{DR3}) we find that the only operation \(R_r^r\) that leaves either node 1, 2 or 3 ( state variables \(x_1,x_2, \text{ or }x_3\)) invariant is just the identity operation \(E\), and it is straightforward to see that (\ref{Thm6}) \( \neq0\) for all choices of \(p,i \text{ and }r\) since there is only one group operation that leaves the \(r\)th state variable invariant, \(R_r^r=E\), for \(r=1,2,3\). Thus, motif 7 cannot be NCS or NOS and must be controllable and observable from any node. \textit{Corollary 1} to \textit{Theorem 6} from \cite{Rubin1972} contains and expands this result directly to any network with only rotational symmetry (i.e. \(C_n\) groups), with the caveat that a network with a state variable that is invariant under all the group operations (motif 7 doesn't have such a state variable) will be NCS and NOS if the input and output are coupled to that variable. 

These representational group theoretic results explain our nonlinear results in section \ref{results}, and clearly demonstrate that different types of symmetry have different effects on the controllability and observability of the networks containing them. While we explicitly assume system matrices with zeros on the diagonal (for simplicity of the calculations) these results hold with generic entries on the diagonal as long as those entries are chosen to preserve the symmetry (e.g. the system matrix \(A\) for motif 1 and 7 has \(a_{11}=a_{22}=a_{33}\) and motif 3 has \(a_{11}=a_{33}\), not shown). Linearization of the system equations in (\ref{FNsys}) would result in a system matrix \(A\) with a non-zero diagonal \cite{Cowan2012}, and is typically done in the analysis of nonlinear networks \cite{Pecora2014} when utilizing such linear analysis techniques. Our computational results demonstrate the utility of this approach in providing insight into the controllability and observability of complex nonlinear networks that have not been linearized.

\subsection{Application to Structurally Controllability (Observability)}
It is interesting to note that the demonstration of our results above and those in \cite{Rubin1972} complement and expand Lin's seminal theorems on structural controllability \cite{Lin1974}. Essentially, a network with system matrix \(A\) and input function \(B\) (the pair \((A,B)\)) are assumed to have two types of entries, non-zero generic entries, and fixed entries which are zero. The position of the zero entries leads to the notion of the structure of the system, where different systems with zeros in the same locations are considered structurally equivalent. With this definition of structure, we arrive at the definition for structurally controllability which states that a pair \((A',B')\) is structurally controllable if and only if there exists a controllable pair \((A'',B'')\) with the same structure as \((A',B')\). The major assumption of this work is that a system deemed to be structurally controllable could indeed be uncontrollable due to the specific entries in \(A\) and \(B\), which for a practical application are assumed to be uncertain estimates of the system parameters and thus subject to modification. 
While Lin's theorems did not explicitly cover symmetry, any network pair \((A,B\) containing symmetry implies constraints on the non-zero entries in \((A,B)\), which is necessary to guarantee that symmetry is present. Thus considering only \cite{Lin1974}, a network with symmetry could be structurally controllable (observable \cite{Rech1990}) as long as the graph of the system contains no dilations\footnote{Defined in the appendix.} or isolated nodes, but NCS (NOS) due to the symmetry. These two theorems together paint a more complete picture of controllability (observability) than either alone as shown in section \ref{results} and \ref{SymGroupTheory}, where both are used in concert to explain and understand why certain network motifs were not controllable or observable from particular nodes. Structural controllability (observability) is a more general result, as it does not depend on the explicit non-zero entries of the system pair \((A,B)\) (necessary, but not sufficient), while a network that has the NCS (NOS) property is due to specific sets of the non-zero entries in \((A,B)\) that define the symmetry contained by the system. 

Additionally, \cite{Lin1974} defined two structures called a ``stem" (our motif 8 controlled from node 3) and a``bud" (our motif 7 controlled from any node) which are always structurally controllable. While both are easily shown to be structurally controllable \cite{Lin1974}, including \textit{Theorem 6} and its \textit{Corollary 1} from \cite{Rubin1972} we can take this a step further and declare that any ``bud" network (of arbitrary size) containing only rotations is not only structurally controllable, but also fully controllable (or never NCS). The dual of these structures for observability is also defined in \cite{Rech1990}, and \textit{Theorem 12} and its \textit{Corollary 1} from \cite{Rubin1972} completes the statement in a similar fashion for observability. Since networks containing only rotation groups or ``buds" in Lin's terminology are always controllable, we see that in some cases, symmetries alone will not destroy the controllability of structurally controllable networks.

\section{Discussion}
Despite the growing importance of exploring observability and controllability in complex graph directed networks, there has been little exploration of nonlinear networks with explicit symmetries.  We here report, to our knowledge, the first exploration of symmetries in nonlinear networks, and show that observability and controllability are a function of the specific type of symmetry, the spatial location of nodes sampled or controlled, the strength of the coupling, and the time evolution of the system.
 
In networks with structural symmetries, group representation theory provides deep insights into how the specific set of symmetry operations possessed by a network will influence its observability and controllability, and can aid in controller or observer design by obtaining a modal decomposition of the network equations into decoupled controllable and uncontrollable (observable and unobservable) subspaces. This knowledge will permit the intelligent placement of the minimum number of sensors and actuators that render a system containing symmetry fully controllable and observable. Additionally, breaking symmetry through randomly altering the coupling strengths established substantial observability or controllability that was absent in the fully symmetric case. In cases where increasing the overall level of coupling strength decreased the observability (controllability), such strong coupling eventually pushed the system towards or through a reverse Hopf bifurcation from limit cycle to a stable equilibrium point, where the lack of dynamic movement of the system then severely decreased the observability (controllability). Intuitively this results from the Lie derivatives (brackets) becoming small as the rate of change of the system trajectories goes to zero. The sensitivity of observability and controllability to the trajectories taken through phase space implies that the choice of control input to a system has to be selected carefully as a poor choice could drive the system into a region that has little to no controllability or observability, thereby thwarting further control effort and/or causing observation of the full system to be lost or limited. Furthermore, when using an observer model for observation or control the regions of local high observability could be utilized to optimize the coupling of the model to a real system by only estimating the full system state when the system transverses observable regions of phase space. 

Observation (control) in motifs 2, 3, 4, 5 and 6 suggests a relationship between the degree of connections into and out of a node and its effective observability (controllability). In general, the more direct connections into an observed node, the higher the observability from that node, and the duality suggests that the more direct number of outgoing connections from a controlled node leads to higher controllability than from other less connected nodes. The high degree `hub' nodes were not the most effective driver nodes in complex networks using linear theory \cite{Liu2011}, and extending nonlinear results to more complex networks with symmetries is a challenge for future work, which may benefit from linear analysis of the connection topology utilizing group representation theory.

When observing kinematics and dynamics of rigid body mechanics obeying Newton's laws with \(SE(3)\) group symmetry, such symmetries must be preserved in constructing an observer (controller) \cite{Bonnabel2008}. In the observation of graph directed networks containing transitive networks, one can observe from any point equivalently within such transitive components \cite{Liu2012}. In the control of graph directed networks, the minimum number of control points were related to the maximal matching nodes \cite{Liu2011}. In \cite{Russo2011}, contraction theory was used to determine symmetric synchronous subspaces - these spaces actually correspond to our regions without observability or controllability. In fact, the proof of observability is that initial conditions and trajectories do not contract \cite{Joly2012}. Furthermore, it is clear that the groupoid input equivalence classes (such as our motifs 6 and 7, see figure 21 in \cite{Golubitsky2006}) are not equivalently observable or controllable - note that only 1 node can serve as an observer node in motif 6 regardless of coupling strength (our Figure \ref{ObsLCNS}). Indeed, whether virtual networks \cite{Russo2011} with particular groupoid equivalent symmetries serve as detectors of observability and controllability remains unresolved at this time.

Our deep knowledge of symmetries and observers in classical mechanics \cite{Bonnabel2008} do not readily translate to graph directed networks. Further development of a theory of observability and controllability for nonlinear networks with symmetries is a vital open problem for future work.


\begin{acknowledgments}
Supported by grants from the National Academies - Keck Futures Initiative, NSF grant DMS 1216568, and Collaborative Research in Computational Neuroscience NIH grant 1R01EB014641.
\end{acknowledgments}

\bibliography{ObservabilityAndControllabilityOfNeuronalNetworkMotifsPRX.bib}

\appendix
\section{Supplemental Information}
\subsection{Construction of Differential Embedding Map and Lie Brackets}
As an example case we begin constructing the observability matrix for motif 1 (shown in Figure \ref{M1Chaos}), where the Fitzhugh-Nagmuo (FN) network equations form the nonlinear vector field \(\mathbf{f}\):
\begin{equation}
\label{NLvect}
	\begin{gathered}
		\mathbf{f}
		\left\{
		\begin{array}{l l}
			f_1=c (v_1-\frac{v_1^3}{3}-w_1+\sum_{j=2,3}f_{NL}(v_j,d_{1j}))\\
			f_2=v_1-bw_1+a\\
			f_3=c (v_2-\frac{v_2^3}{3}-w_2+\sum_{j=1,3}f_{NL}(v_j,d_{2j}))\\
			f_4=v_2-bw_2+a\\
			f_5=c (v_3-\frac{v_3^3}{3}-w_3+\sum_{j=1,2}f_{NL}(v_j,d_{3j}))\\
    			f_6=v_3-bw_3+a\\
		\end{array} \right.
	\end{gathered}
\end{equation}
and the measurement function for node 1 in motif 1 is \(y=C\mathbf{x}(t)=[1,0,0,0,0,0]\mathbf{x}(t)=v_1\). We construct the differential embedding map by taking the Lie derivatives (\ref{LieDeriv}) from \(\mathfrak{L}^{0}_{f}(y)\) to \(\mathfrak{L}^{5}_{f}(y)\) as:
\begin{equation}
\label{DEmap}
	\begin{gathered}
	\phi 
	\left\{
		\begin{array}{l l}
			\phi_1=y=v_1\\
			\phi_2=\frac{\partial y}{\partial v_1} \cdot f_1=f_1 \\
			\phi_3=\frac{\partial \phi_2}{\partial v_1}f_1+\frac{\partial \phi_2}{\partial w_1}f_2+\frac{\partial \phi_2}{\partial v_2}f_3+\ldots+\frac{\partial \phi_2}{\partial w_3}f_6\\
			\phi_4=\frac{\partial \phi_3}{\partial v_1}f_1+\frac{\partial \phi_3}{\partial w_1}f_2+\frac{\partial \phi_3}{\partial v_2}f_3+\ldots+\frac{\partial \phi_3}{\partial w_3}f_6\\
			\phi_5=\frac{\partial \phi_4}{\partial v_1}f_1+\frac{\partial \phi_4}{\partial w_1}f_2+\frac{\partial \phi_4}{\partial v_2}f_3+\ldots+\frac{\partial \phi_4}{\partial w_3}f_6\\
    			\phi_6=\frac{\partial \phi_5}{\partial v_1}f_1+\frac{\partial \phi_5}{\partial w_1}f_2+\frac{\partial \phi_5}{\partial v_2}f_3+\ldots+\frac{\partial \phi_5}{\partial w_3}f_6\\
		\end{array} \right.
	\end{gathered}
\end{equation}
where \(\frac{\partial \phi_i}{\partial x_j} \) is the partial derivative of the \(i^{th}\) row of the embedding map \( \phi \), with respect to the \(j^{th}\) state variable. We obtain the observability matrix by taking the Jacobian of (\ref{DEmap}). In this FN network the observability matrix is dependent on the state variables and is thus a function of the location in phase space as the system evolves in time. Letellier et al. \cite{Letellier1998} used averages of the observability index over the state trajectories in phase space as a qualitative measure of observability. We adopt this convention when computing observability of various network motifs. The indices are computed for each time point in the trajectory, and then the average is taken over all of the trajectories.

Constructing the nonlinear controllability matrix for motif 1 from node 1 begins with the control input function \(\mathbf{g}=B\mathbf{u}(t)=[1,0,0,0,0,0]^T\) and its Lie bracket with respect to the nonlinear vector field \(\mathbf{f}\) in (\ref{NLvect}). We exclude the internal driving square wave function here since it is connected to all three nodes, would provide no contribution in the Lie bracket mapping, and we are interested in the mapping from the control input \(\mathbf{g}\) to the states in order to determine if the system can be controlled, 
\begin{equation}
\label{LieBracket2}
	\begin{gathered}
		\begin{bmatrix}\mathbf{f},\mathbf{g}\end{bmatrix}=\underbrace{\frac{\partial\mathbf{g}}{\partial\mathbf{x}}\mathbf{f}}_{0}-\frac{\partial\mathbf{f}}{\partial\mathbf{x}}\mathbf{g} =
		\left\{
		\begin{array}{l l}
			-\frac{\partial f_1}{\partial v_1}g_1-\ldots-\frac{\partial f_1}{\partial w_3}g_6\\
			-\frac{\partial f_2}{\partial v_1}g_1-\ldots-\frac{\partial f_2}{\partial w_3}g_6\\
			-\frac{\partial f_3}{\partial v_1}g_1-\ldots-\frac{\partial f_3}{\partial w_3}g_6\\
			-\frac{\partial f_4}{\partial v_1}g_1-\ldots-\frac{\partial f_4}{\partial w_3}g_6\\
			-\frac{\partial f_5}{\partial v_1}g_1-\ldots-\frac{\partial f_5}{\partial w_3}g_6\\
			-\frac{\partial f_6}{\partial v_1}g_1-\ldots-\frac{\partial f_6}{\partial w_3}g_6\\
		\end{array} \right.
	\end{gathered}
\end{equation}
where \( \frac{\partial\mathbf{g}}{\partial\mathbf{x}}=0 \) since \(\mathbf{g}\) is the same at each node, \(\frac{\partial f_i}{\partial x_j} \) is the partial derivative of the \(i^{th}\) row of the nonlinear vector field \( \mathbf{f}(x) \) with respect to the \(j^{th}\) state variable, and \(g_i\) is the \(i^{th}\) component of the input vector \(\mathbf{g}\). We construct the controllability matrix from the definitions in equations (\ref{LieBracket} and \ref{CtrlbMatrixNL}), as the control input function \(\mathbf{g}\) and its higher Lie Brackets from \((ad^{1}_{\mathbf{f}},\mathbf{g})\) to \((ad^{5}_{\mathbf{f}},\mathbf{g})\) with respect to the nonlinear vector field system equations,
\begin{equation}
\label{CtrlbMatrixNL2}
	Q=
	\begin{bmatrix}
		\mathbf{g},[\mathbf{f},\mathbf{g}],[\mathbf{f},[\mathbf{f},\mathbf{g}]],(ad^{3}_{\mathbf{f}},\mathbf{g}),(ad^{4}_{\mathbf{f}},\mathbf{g}),(ad^{5}_{\mathbf{f}},\mathbf{g})
	\end{bmatrix}
\end{equation}

\subsection{Observability and Controllability Index Distribution}
\begin{figure}[!]
\centering
\includegraphics[width=3.4in, clip=true, trim = 23 2 53 4]{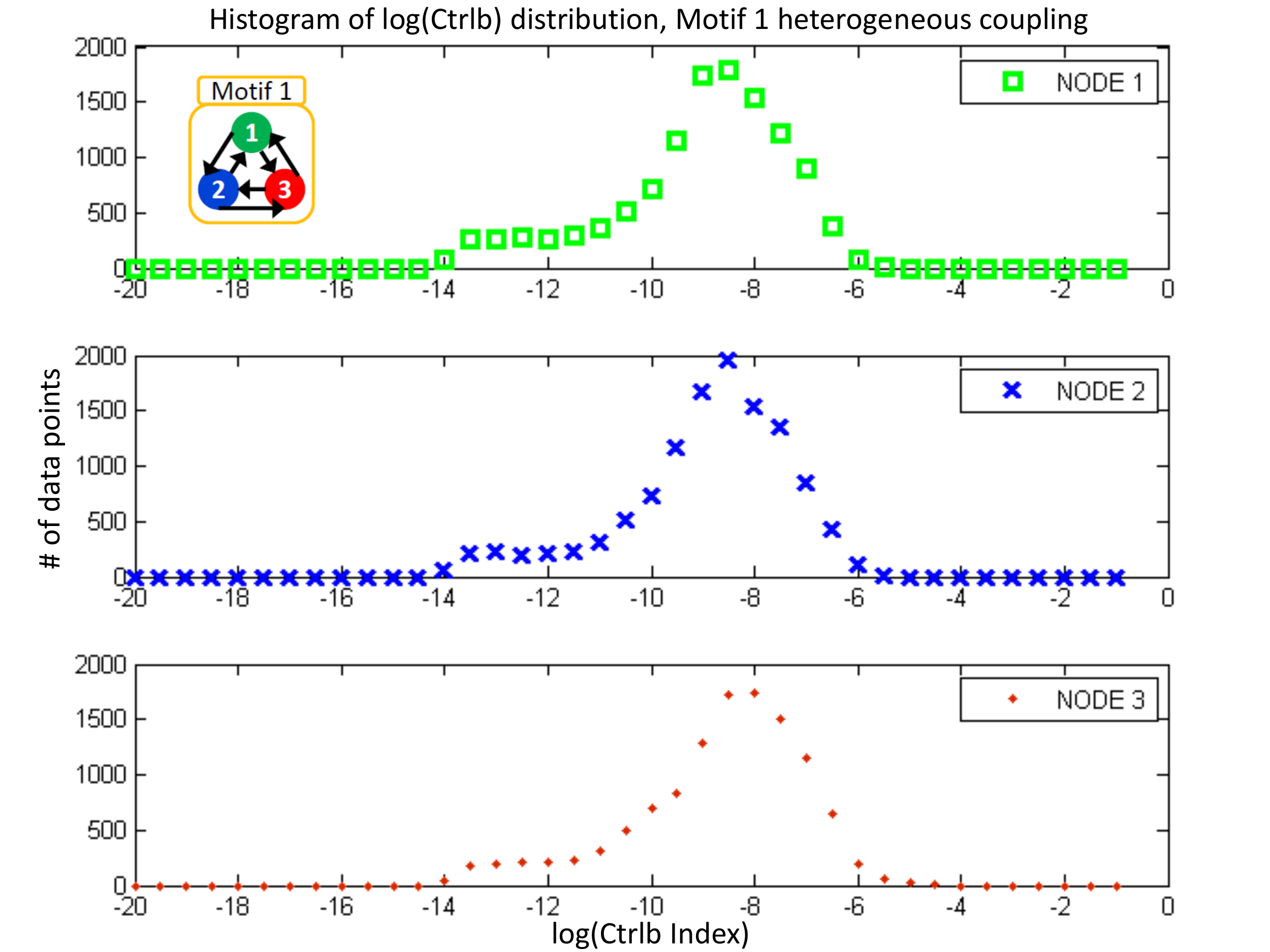}
\caption{\label{histogram}The histogram of the log-scaled controllability indices for motif 1 with heterogeneous coupling and chaotic dynamics}
\end{figure}

Log-scaled histograms (Figure \ref{histogram}) of the index distributions reveal that the local observability (controllability) along the trajectories in phase space are close to a log-normal distribution. After removing zeros from the data, these log-normal distribution fits were computed and verified with the \(\chi ^2\) test metric for all of the observability and controllability computation cases that contained an adequate number of data points to accurately compute the fit (over \(90\%\) of the data). The \(\chi ^2\) test for goodness of fit confirmed that the data come from a log-normal distribution with 95\% confidence. This type of zeros-censored log-normal distribution is known as a delta distribution \cite{Aitchison1957}, and the estimated mean \(\kappa\) and variance \(\rho^2\) are adjusted to account for the proportion of data points that are zero, \(\delta\), as follows
\begin{equation}
\label{adjMeanandVar}
	\begin{aligned}
		\delta &= \frac{\#\{i : x_i=0\}}{n} \\
		\kappa &= (1-\delta) e^{\mu+0.5\sigma^2} \\
		\rho^2 &= (1-\delta) e^{2\mu+\sigma^2} (e^{\sigma^2}-(1-\delta)),
	\end{aligned} 
\end{equation}
where \(\mu\) and \(\sigma\) are the mean and variance associated with the lognormal distribution computed from the non-zero data. We use these equations to compute the statistics in the plots in the results section (Figures \ref{M1Chaos} to \ref{3DChaos}).

\subsection{Group Representation Analysis of Symmetries in Motif 1}
We examine motif 1 in Figure \ref{symFig} which has \(\mathbf{S_3}\) symmetry. Determined from (\ref{char}), there are 3 classes of group elements \(\mathscr{C}_1=\{E\},\mathscr{C}_2=\{\sigma_1,\sigma_2,\sigma_3\} \text{ and }\mathscr{C}_3=\{C_3,C_3^2\}\). Reduction of \(D(R)\) yields the two, 1-dimensional and one 2-dimensional (\(l_1=l_2=1, l_3=2\)) irreducible representations (computed from (\ref{dim})) \(D^{(1)}(R), D^{(2)}(R) \text{ and }D^{(3)}(R)\) of \(\mathbf{S_3}\), which are found in Table \ref{IrrRep} and from (\ref{ap}) appear 1, 0 and 2 times in \(D(R)\) respectively. Forming the generating matrix in equation (\ref{G}) we construct \(\alpha\) for motif 1 as follows
\begin{equation}
\renewcommand{\arraystretch}{1.0}
\begin{aligned}
	G_{1}^{(1)}&=\sum_{R\in S_3} D^{(1)}(R)_{1 1}^{*}D(R)I \\
	&=1\begin{bmatrix}
		1 \!\!& 0 \!\!& 0 \\
		0 \!\!& 1 \!\!& 0 \\
		0 \!\!& 0 \!\!& 1 
	\end{bmatrix}\!\! +
	1\begin{bmatrix}
		1 \!\!& 0 \!\!& 0 \\
		0 \!\!& 0 \!\!& 1 \\
		0 \!\!& 1 \!\!& 0 
	\end{bmatrix}\!\! +
	1\begin{bmatrix}
		0 \!\!& 0 \!\!& 1 \\
		0 \!\!& 1 \!\!& 0 \\
		1 \!\!& 0 \!\!& 0 
	\end{bmatrix}\!\! +\hdots \\
	&\quad\,\,1\begin{bmatrix}
		0 \!\!& 1 \!\!& 0 \\
		1 \!\!& 0 \!\!& 0 \\
		0 \!\!& 0 \!\!& 1 
	\end{bmatrix}\!\! +
	1\begin{bmatrix}
		0 \!\!& 1 \!\!& 0 \\
		0 \!\!& 0 \!\!& 1 \\
		1 \!\!& 0 \!\!& 0 
	\end{bmatrix}\!\! +
	1\begin{bmatrix}
		0 \!\!& 0 \!\!& 1 \\
		1 \!\!& 0 \!\!& 0 \\
		0 \!\!& 1 \!\!& 0 
	\end{bmatrix} \\
	&=\;\;\,\begin{bmatrix}
		2 \!\!& 2 \!\!& 2 \\
		2 \!\!& 2 \!\!& 2 \\
		2 \!\!& 2 \!\!& 2 
	\end{bmatrix},
\end{aligned}
\end{equation}
where each linearly independent row of \(G\) is a column of \(\alpha\), and thus
\begin{equation}
\renewcommand{\arraystretch}{1.0}
	\begin{bmatrix}
		2 \\ 2 \\ 2
	\end{bmatrix}
	\xrightarrow[\text{normalize}]{ }
	\begin{bmatrix}
		\frac{1}{\sqrt{3}} \\ \frac{1}{\sqrt{3}} \\ \frac{1}{\sqrt{3}}
	\end{bmatrix}
	=\begin{bmatrix}
		\alpha_{11} \\ \alpha_{12} \\ \alpha_{13}
	\end{bmatrix}
\end{equation}
defines the first column of \(\alpha\). We know from (\ref{ap}) that \(D^{(2)}(R)\) appears zero times in \(D(R)\) and thus yields no contribution to \(\alpha\). Continuing, we have the last two computations from the 2-dimensional irreducible representation \(D^{(3)}\) (one for each row)
\begin{equation}
\renewcommand{\arraystretch}{1.0}
\begin{aligned}
	G_{1}^{(3)}&=\sum_{R\in S_3} D^{(3)}(R)_{1 1}^{*}D(R)I \\
	&=1\begin{bmatrix}
		1 \!\!& 0 \!\!& 0 \\
		0 \!\!& 1 \!\!& 0 \\
		0 \!\!& 0 \!\!& 1 
	\end{bmatrix}\!\! -
	1\begin{bmatrix}
		1 \!\!& 0 \!\!& 0 \\
		0 \!\!& 0 \!\!& 1 \\
		0 \!\!& 1 \!\!& 0 
	\end{bmatrix}\!\! +
	\frac{1}{2}\begin{bmatrix}
		0 \!\!& 0 \!\!& 1 \\
		0 \!\!& 1 \!\!& 0 \\
		1 \!\!& 0 \!\!& 0 
	\end{bmatrix}\!\! +\hdots \\
	&=\;\;\,\begin{bmatrix}
		0 \!\!& 0 \!\!& 0 \\
		0 \!\!& \frac{3}{2} \!\!& -\frac{3}{2} \\
		0 \!\!& -\frac{3}{2} \!\!& \frac{3}{2} 
	\end{bmatrix},
\end{aligned}
\end{equation}
which after normalization yields
\begin{equation}
\renewcommand{\arraystretch}{1.0}
	\begin{bmatrix}
		0 \\ \frac{3}{2} \\ -\frac{3}{2}
	\end{bmatrix}
	\xrightarrow[\text{normalize}]{ }
	\begin{bmatrix}
		0 \\ \frac{1}{\sqrt{2}} \\ -\frac{1}{\sqrt{2}}
	\end{bmatrix}
	=\begin{bmatrix}
		\alpha_{21} \\ \alpha_{22} \\ \alpha_{23}
	\end{bmatrix},
\end{equation}
and
\begin{equation}
\renewcommand{\arraystretch}{1.0}
\begin{aligned}
	G_{2}^{(3)}&=\sum_{R\in S_3} D^{(3)}(R)_{2 2}^{*}D(R)I \\
	&=1\begin{bmatrix}
		1 \!\!& 0 \!\!& 0 \\
		0 \!\!& 1 \!\!& 0 \\
		0 \!\!& 0 \!\!& 1 
	\end{bmatrix}\!\! +
	1\begin{bmatrix}
		1 \!\!& 0 \!\!& 0 \\
		0 \!\!& 0 \!\!& 1 \\
		0 \!\!& 1 \!\!& 0 
	\end{bmatrix}\!\! -
	\frac{1}{2}\begin{bmatrix}
		0 \!\!& 0 \!\!& 1 \\
		0 \!\!& 1 \!\!& 0 \\
		1 \!\!& 0 \!\!& 0 
	\end{bmatrix}\!\! +\hdots \\
	&=\;\;\,\begin{bmatrix}
		2 \!\!& -1 \!\!& -1 \\
		-1 \!\!& \frac{1}{2} \!\!& \frac{1}{2} \\
		-1 \!\!& \frac{1}{2} \!\!& \frac{1}{2} 
	\end{bmatrix}
\end{aligned}
\end{equation}
yields the last column of \(\alpha\) (after normalization)
\begin{equation}
\renewcommand{\arraystretch}{1.0}
	\begin{bmatrix}
		2 \\ -1 \\ -1
	\end{bmatrix}
	\xrightarrow[\text{normalize}]{ }
	\begin{bmatrix}
		2 \\ -\frac{1}{\sqrt{6}} \\ -\frac{1}{\sqrt{6}}
	\end{bmatrix}
	=\begin{bmatrix}
		\alpha_{31} \\ \alpha_{32} \\ \alpha_{33}
	\end{bmatrix}.
\end{equation}
Finally, the coordinate transformation matrix \(\alpha\) is
\begin{equation}
\renewcommand{\arraystretch}{1.0}
	\alpha=
	\begin{bmatrix}
		\frac{1}{\sqrt{3}} & \frac{2}{\sqrt{6}} & 0 \\
		\frac{1}{\sqrt{3}} & -\frac{1}{\sqrt{6}} & \frac{1}{\sqrt{2}} \\
		\frac{1}{\sqrt{3}} & -\frac{1}{\sqrt{6}} & -\frac{1}{\sqrt{2}}
	\end{bmatrix}.
\end{equation}
and the computation is concluded in section \ref{ConstrAlpha}.

\renewcommand{\arraystretch}{1.5}
\begin{table*}[!th]
\caption{\label{IrrRep}Irreducible representations for \(S_3\) symmetry.}
\begin{ruledtabular}
\begin{tabular}{ c | c c c c c c }
	R & E & $\sigma_1$ & $\sigma_2$ & $\sigma_3$ & $C_3$ & $C_3^2$ \\\hline
	$D^{(1)}(R)$ & 1 & 1 & 1 & 1 & 1 & 1 \\
	$D^{(2)}(R)$ &1 &-1 &-1 &-1 & 1 & 1 \\
	$D^{(3)}(R)$ & $\begin{bmatrix}1 & 0\\ 0 & 1\end{bmatrix}$ & $\begin{bmatrix}-1 & 0\\ 0 & 1\end{bmatrix}$ & $\begin{bmatrix}\frac{1}{2} & -\frac{\sqrt{3}}{2}\\ -\frac{\sqrt{3}}{2} & -\frac{1}{2}\end{bmatrix}$ & $\begin{bmatrix}\frac{1}{2} & \frac{\sqrt{3}}{2}\\ \frac{\sqrt{3}}{2} & -\frac{1}{2}\end{bmatrix}$ & $\begin{bmatrix}-\frac{1}{2} & -\frac{\sqrt{3}}{2}\\ \frac{\sqrt{3}}{2} & -\frac{1}{2}\end{bmatrix}$ & $\begin{bmatrix}-\frac{1}{2} & \frac{\sqrt{3}}{2}\\ -\frac{\sqrt{3}}{2} & -\frac{1}{2}\end{bmatrix}$
\end{tabular}
\end{ruledtabular}
\end{table*}

\subsection{Dilations of the graph of (A,B)}
In \cite{Lin1974}, the graph \(G\) of the pair \((A,B)\) is defined as a graph of \(n+1\) nodes \(e_1,e_2, \hdots, e_{n+1}\), where \(n\) is the dimension of \(A\), and \(e_{n+1}\) is called the ``origin" (the input). The vertex set \(S=\{e_1,e_2, \hdots,e_n\}\) is defined as the set of all nodes in \(G\) excluding the origin (\(e_{n+1}\)). A dilation is present in \(G\) if and only if \(|T(S)|<|S|\), where \(T(S)\) is defined as the set of all nodes that have a directed edge pointing to a node in the set \(S\).

\end{document}